\documentclass[american,prl,reprint]{revtex4-2}
\usepackage[T1]{fontenc}
\usepackage[latin9]{inputenc}
\usepackage{color}
\usepackage{babel}
\usepackage{mathtools}
\usepackage{amsmath}
\usepackage{amssymb}
\usepackage{graphicx}
\usepackage[unicode=true,pdfusetitle,
 bookmarks=true,bookmarksnumbered=false,bookmarksopen=false,
 breaklinks=true,pdfborder={0 0 1},backref=false,colorlinks=true]
 {hyperref}
\hypersetup{
 pdfborderstyle=,citecolor=blue}

\makeatletter
\usepackage{babel}
\usepackage{array}
\usepackage{multirow}
\usepackage{times}
\usepackage{longtable} 

\makeatother

\begin{document}

\title{Fluctuation-Response Design Rules for Nonequilibrium Flows}
\author{Ying-Jen Yang}
\email{ying-jen.yang@stonybrook.edu}
\affiliation{Laufer Center of Physical and Quantitative Biology, Stony Brook University}
\author{Ken A. Dill}
\affiliation{Laufer Center of Physical and Quantitative Biology, Stony Brook University}
\affiliation{Department of Physics, Stony Brook University}
\affiliation{Department of Chemistry, Stony Brook University}

\begin{abstract}
Biological machines like molecular motors and enzymes operate in dynamic cycles representable as stochastic flows on networks.  Current stochastic dynamics describes such flows on fixed networks.  Here, we develop a scalable approach to \textit{network design} in which local transition rates can be systematically varied to achieve global dynamical objectives. It is based on the fluctuation-response duality in the recent Caliber Force Theory---a path-entropy variational formalism for nonequilibria.  This approach scales efficiently with network complexity and gives new insights, for example revealing the transition from timing- to branching-dominated fluctuations in a kinesin motor model.
\end{abstract}
\maketitle
\textbf{Introduction:} 
Consider a stochastic dynamical process transitioning among a network of states.  In biophysics alone, such stochastic network flows are involved in the molecular mechanisms of motors and pumps \cite{kolomeisky_molecular_2007,liepelt_kinesins_2007,wagoner_mechanisms_2019}, ultrasensitive switches in flagella \cite{duke_conformational_2001,tu_nonequilibrium_2008}, the catalytic and allosteric actions in enzymes \cite{monod_nature_1965,hopfield_kinetic_1974,ninio_kinetic_1975,qian_single-molecule_2002}, phosphorylation cycles \cite{goldbeter_amplified_1981,qian_phosphorylation_2007}, energy and chemical transduction \cite{hill_free_1989,brown_theory_2020} and others. A fundamental goal is to control the long-term statistics of flows, given the local transition rates.\\

While Markov State Models and Master Equations describe these dynamics \cite{schnakenberg_network_1976}, they do not prescribe design. What is missing are principles for the optimization, control, and evolution of network flows: how should one systematically tune local transition rates to achieve specific global functions? Mathematically, this requires solving the response gradient---knowing exactly how local perturbations shift the long-term stochastic dynamics. However, raw gradients evaluated case-by-case do not reveal general principles. To navigate systematically, we exploit a key physical connection: general response relations are encoded in the system's fluctuations.\\

Echoing Onsager's regression hypothesis, recent nonequilibrium (NEQ) advances have established diverse fluctuation-response relations---equalities~\cite{baiesi_fluctuations_2009,seifert_fluctuation-dissipation_2010,owen_universal_2020,aslyamov_nonequilibrium_2024,zheng_unified_2025,aslyamov_nonequilibrium_2025} or bounds~\cite{owen_universal_2020,fernandes_martins_topologically_2023,owen_size_2023,ptaszynski_dissipation_2024,aslyamov_nonequilibrium_2025}---revealing that ``noisier'' mechanisms are inherently more susceptible to perturbative control. 
However, operationalizing this for NEQ flow design has some obstacles. 
First, design demands gradients, not just bounds. 
While performance limits are insightful \cite{owen_universal_2020,fernandes_martins_topologically_2023,owen_size_2023,ptaszynski_dissipation_2024,aslyamov_nonequilibrium_2025}, inequalities inherently lack the local directional information needed to navigate parameter space. Second, we need to design and control the flux statistics too, not just the average state occupancy or the mean fluxes. 
Response relations on state occupancy~\cite{baiesi_fluctuations_2009,owen_size_2023} do not control flux fluctuations. Third, scalability is paramount. 
Standard techniques, such as differentiating generating functions~\cite{koza_general_1999,barato_thermodynamic_2015} or summing over spanning trees~\cite{hill_free_1989,schnakenberg_network_1976,owen_universal_2020,fernandes_martins_topologically_2023}, offer physical insights but scale unfavorably with system size, rendering large-network optimization computationally demanding. 
We need an actionable, complete and scalable framework.  \\

Here, we address this need by establishing a principle-based design framework rooted in Caliber Force Theory (CFT)~\cite{yang_principled_2025}. By projecting the independent transition noise~\cite{zheng_unified_2025} onto the orthogonal counting observables in CFT, we show that kinetic fluctuation-response relations originate from the theory's fundamental observable-force conjugacy.
This structure is encoded in a Jacobian matrix linking transition rates to \textit{forces}.
This underpinning provides a systematic and scalable design framework.
Applied to a data-fitted kinesin model \cite{liepelt_kinesins_2007}, it parses fluctuations to reveal the chemo-mechanical covariances underlying a load-dependent noise transition.
Computationally, our framework overcomes the scalability bottleneck in evaluating gradients for the means and variances of fluxes in large networks.
Analytically, it unifies NEQ response relations, extending previous results \cite{owen_universal_2020,aslyamov_nonequilibrium_2024, aslyamov_nonequilibrium_2025} and deriving kinetic bounds reflecting population depletion and Le Chatelier-like compensation.
This framework transforms the geometry of spontaneous fluctuations into a scalable ``road-map'' for gradient-based design.\\

\textbf{Theoretical Framework}: 
Our foundation here is CFT, a dynamical variational formalism that mirrors the structure of equilibrium (EQ) thermodynamics \cite{yang_principled_2025}. We begin by considering the \textit{path entropy} of Markov jump processes---defined as the logarithm of path probability ratio $\ln (\mathcal{P}_{\textbf{k}} / \mathcal{P}_{\textbf{u}})$. It measures the path-wise entropic cost of driving a unit-rate reference process $\textbf{u}$ to the dynamics with transition rate $\textbf{k}$. For a path $\omega_t$ with duration $t$, it can be expressed in terms of \textit{extensive counting observables} \cite{SM}:
\begin{align}
    \ln \frac{\mathcal{P}_{\textbf{k}}(\omega_t)}{\mathcal{P}_\textbf{u}(\omega_t)} = \sum_{i\neq j} N_{ij}(\omega_t)~\ln k_{ij} - \sum_i T_i(\omega_t) ~ \varepsilon_i (\textbf{k}) ,
    \label{eq:path_entropy_raw}
\end{align}
where $N_{ij}$ is the number of transitions $i\to j$, $T_i$ is the total dwell time of state $i$, and $\varepsilon_i=\sum_{j(\neq i)}(k_{ij}-1)$ is the excess escape rate relative to the unit-rate reference. However, these raw observables have undesirable redundancies. Conservation laws---time additivity $\sum_i T_i = t$ and Kirchhoff's current balance $\sum_{j(\neq i)} N_{ij} \sim \sum_{j (\neq i)} N_{ji}$ \footnote{We use standard asymptotic analysis notation $a \sim b$ to represent asymptotic equivalency, $\lim_{t\rightarrow \infty} a(t)/b(t) =1$.}---enforce strong coupling among them in the long term, rendering the naive counts $(N_{ij}, T_i)$ degenerate axes for control. \\


CFT resolves this redundancy by identifying the \textit{asymptotically orthogonal} counting variables $\textbf{X}$ and relating the path entropy with their conjugate forces  $\boldsymbol{\mathfrak{F}}$---defined as the path entropy derivatives~\cite{yang_principled_2025,SM}:
\begin{equation}
    \ln \frac{\mathcal{P}_{\textbf{k}}(\omega_t)}{\mathcal{P}_\textbf{u}(\omega_t)} \sim \boldsymbol{\mathfrak{F}}(\textbf{k}) \cdot \textbf{X}(\omega_t) - \mathfrak{c}(\textbf{k})~t
    \label{eq:cft_form}
\end{equation}
where the caliber $(\mathfrak{c}~t)$ resembles for our NEQ processes the role that the free energy plays for EQ processes.  
The vector $\textbf{X} = (\Phi_{ij}, T_{n}, \Psi_{c})$ is a complete observable basis consistent with the asymptotic constraints:
(i) \textbf{edge traffic} $\Phi_{ij} = N_{ij} + N_{ji}$;
(ii) \textbf{dwell times} $T_{n}$ (excluding a reference $m$); and
(iii) \textbf{cycle net fluxes} $\Psi_{c} = N_{ab} - N_{ba}$, defined by the net flux across the chord $ab$ of a fundamental cycle $c$ (mnemonic: the time irreversible counterpart to traffic, $\Psi \equiv \text{``i''} + \Phi$).
Dividing by time gives their rates $\textbf{x}=(\phi_{ij},f_n,\psi_c)$,  and taking steady-state averages yields the intensive rate coordinate for the parameter space $\langle \textbf{x} \rangle = \langle \textbf{X} \rangle/t=(\tau_{ij},\pi_n,J_c)$.
Their conjugate forces $\boldsymbol{\mathfrak{F}}$ are the affinities to edge exchange, node dwelling, and cycle completion \cite{yang_principled_2025}:
\begin{align}
\mathfrak{F}_{\text{edge},ij} & =\frac{1}{2}\ln{k_{ij}k_{ji}},\nonumber \\
\mathfrak{F}_{\text{node},n} & =\sum_{i(\neq m)}(k_{mi}-1)-\sum_{j(\neq n)}(k_{nj}-1),\nonumber \\
\mathfrak{F}_{\text{cycle},c} & =\frac{1}{2}\ln{\frac{k_{i_{0}i_{1}}k_{i_{1}i_{2}}\cdots k_{i_{\sigma}i_{0}}}{k_{i_{0}i_{\sigma}}k_{i_{\sigma}i_{\sigma-1}}\cdots k_{i_{1}i_{0}}}}\label{eqs: force-expression-in-terms-of-the-rates}
\end{align}
where $i_{0}i_{1}i_{2}...i_{\sigma}i_{0}$ is the state sequence of
the cycle $c$.
Crucially, the caliber rate  $\mathfrak{c}(\textbf{k}) = \sum_{j\neq m} (k_{mj} - 1)$ functions as the log dynamic partition function that generates observable statistics when parameterized by the forces $\boldsymbol{\mathfrak{F}}$ \cite{yang_principled_2025,SM}.\\

The asymptotic form of Eq.~\eqref{eq:cft_form} establishes fundamental \textit{conjugate relations} between the observables and the forces.  This conjugate structure dictates a fundamental duality between \textit{observable fluctuations} and their corresponding \textit{force responses}: the average susceptibility of the CFT rate observables $x_\alpha$ to a force $\mathfrak{F}_\beta$---while holding all other forces fixed---is exactly the asymptotic covariance \cite{yang_principled_2025}:
\begin{equation}
    \left. \frac{\partial \langle x_\alpha \rangle}{\partial \mathfrak{F}_\beta} \right|_{\mathfrak{F}_{\gamma \neq \beta}} = \frac{\partial^2 \mathfrak{c}(\boldsymbol{\mathfrak{F})}}{\partial \mathfrak{F}_\alpha ~\partial \mathfrak{F}_\beta} = \lim_{t\rightarrow \infty} t~\mathrm{Cov}[x_\alpha, x_\beta]. 
    \label{eq:cft_fdt}
\end{equation}
Thus, fluctuations equal susceptibilities: (a) the strictly positive variance implies \textit{monotonic} force responses: just as heat capacity is positive, any average rate observable $\langle x_\alpha \rangle\in (\tau_{ij},\pi_n,J_c)$ must increase with its own conjugate drive $\mathfrak{F}_\alpha$; (b) the symmetry of covariance enforces a \textit{generalized Maxwell-Onsager reciprocity} far from equilibrium: the susceptibility of observable $\langle x_\alpha \rangle$ to force $\mathfrak{F}_\beta$ is identical to the response of $\langle x_\beta \rangle$ to force $\mathfrak{F}_\alpha$. \\

\begin{figure}
\begin{centering}
\includegraphics[width=1\columnwidth]{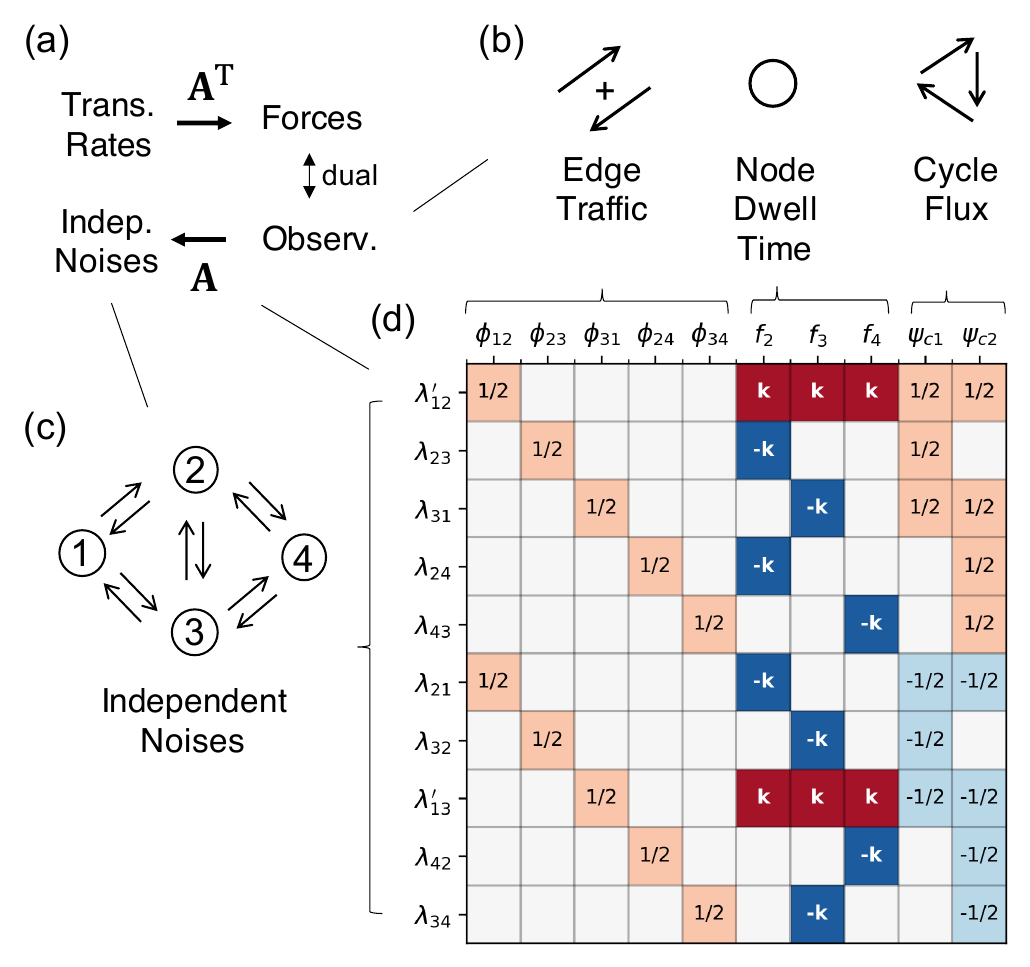}
\par\end{centering}
\caption{\textbf{The sparse structure of fluctuation-response duality.} 
\textbf{(a)} The Jacobian matrix $\textbf{A}$ acts as the bridge between stochasticity and control: it maps physical observables $\textbf{x}$ to independent noise sources $\boldsymbol{\lambda}$ (Fluctuation Geometry), while simultaneously linking transition rates $\ln \textbf{k}$ to conjugate forces $\boldsymbol{\mathfrak{F}}$ (Response Geometry).
\textbf{(b)} The complete observable basis: edge traffic $\phi_{ij}$, node frequency $f_n$, and cycle net flux $\psi_c$.
\textbf{(c)} A representative 4-state network showing that every transition edge generates an intrinsic independent noise source $\lambda_{ij}$.
\textbf{(d)} The explicit structure of the Jacobian $\textbf{A}$ for the 4-state example. 
The symbol $k$ (or $-k$) denotes the transition rate $k_{ij}$ corresponding to the specific row index. The prime notation $\lambda'_{mj}$ denotes the shifted noise source $\lambda'_{mj} = \lambda_{mj} + k_{mj}$ for transitions leaving the reference state $m=1$.
White squares denote structural zeros.
\label{fig:Cartoon-for-three}}
\end{figure}

While the forces $\boldsymbol{\mathfrak{F}}$ have clear physical meanings as affinities, practical control more often involves tuning specific transition rates $k_{ij}$ while fixing others, rather than manipulating one force while clamping the rest.
We thus map the force-response conjugacy in CFT to the rate-response landscape:
\begin{equation}
    \left. \frac{\partial \langle \textbf{x} \rangle}{\partial \ln k_{ij}} \right|_{k_{ab},~ab\neq ij} = \sum_{\beta} \frac{\partial \langle \textbf{x} \rangle}{\partial \mathfrak{F}_\beta}
    \frac{\partial \mathfrak{F}_\beta}{\partial \ln k_{ij}}
    \label{eq:chain_rule}
\end{equation}
The projection is encoded by a sparse matrix $\textbf{A}_{(ij),\beta} \equiv \partial \mathfrak{F}_\beta / \partial \ln k_{ij}$,  the Jacobian linking the transition rates $\textbf{k}$ to the forces $\boldsymbol{\mathfrak{F}}$~\cite{SM}, illustrated in Fig. \ref{fig:Cartoon-for-three}. \\

To further operationalize this, we identify the independent noise sources, $\lambda_{ij} = (N_{ij} - k_{ij}T_i)/t$, in a Markov jump process, whose asymptotic covariances are diagonal~\cite{zheng_unified_2025}:
\begin{equation}
    t\cdot{\rm Cov}[\lambda_{ij},\lambda_{mn}]\sim \pi_{i}k_{ij}~ \delta_{i,m}~ \delta_{j,n}.
    \label{eq:cov_lambda}
\end{equation}
These noise sources are asymptotically linearly spanned by the CFT observables via the same Jacobian $\textbf{A}$~\cite{SM}:
\begin{equation}
    \boldsymbol{\lambda}[\omega_t] \sim \textbf{A} (\textbf{k})~ \textbf{x}[\omega_t] - \nabla_{\ln \textbf{k}} \mathfrak{c}(\textbf{k}).
    \label{eq:linear_span}
\end{equation}
Combining these yields the central \textbf{Response-Inverse-Matrix (RIM)} relation~\cite{SM}:
\begin{align}
    \frac{\partial\left\langle x_\alpha \right\rangle }{\partial \ln k_{ij}} 
    &= \lim_{t\to\infty} {t \cdot \mathrm{Cov}[x_\alpha, \lambda_{ij}]} \nonumber \\
    &= \pi_i k_{ij}~ [\textbf{A}^{-1}]_{\alpha,(ij)} 
    \label{eq:RIM}
\end{align}
where $\textbf{A}^{-1}=\nabla_{\boldsymbol{\mathfrak{F}}} ~\ln \textbf{k}({\boldsymbol{\mathfrak{F}}})$ is the Jacobian from the forces to the transition rates.\\

These equations constitute a \textit{fluctuation-response duality} for rate perturbations, originating from the underlying \textit{observable-force conjugacy}: the Jacobian $\textbf{A}$ maps the CFT conjugates $(\textbf{x}, \boldsymbol{\mathfrak{F}})$ onto the kinetic variables $(\boldsymbol{\lambda}, \ln \textbf{k})$.
While the first equality in Eq. \eqref{eq:RIM} recovers a known result recently highlighted by Zheng and Lu~\cite{zheng_unified_2025}, the second equality reveals the underlying dual geometry, identifying responses explicitly as matrix elements of the inverse Jacobian $\textbf{A}^{-1}$.
This gives three results:
the linear mapping enables parsing fluctuations into independent components (Result 1);
an algebraic closure of derivatives enables scalable gradient evaluation (Result 2);
and Jacobian symmetries yield unified response relations, generalizing recent results \cite{owen_universal_2020,aslyamov_nonequilibrium_2024,aslyamov_nonequilibrium_2025} (Result 3).\\

\textbf{Fluctuation--Response Relations are the key to design and optimization.}  
While the statistical independence of noise sources $\boldsymbol{\lambda}$ is established~\cite{zheng_unified_2025}, it is the linear mapping in Eq.~\eqref{eq:linear_span} that identifies them as the complete, projectable coordinate axes. We exploit this geometric structure to decompose the total covariance between any two observables $x, x'\in(\phi_{ij},f_n,\psi_c)$ into a sum over their projections onto the orthonormal axes:
\begin{equation}
    \mathrm{Cov}(x, x') 
    \sim \sum_{i,j} \mathrm{Cov}[x, \hat{\lambda}_{ij}] \cdot \mathrm{Cov}[x', \hat{\lambda}_{ij}],
    \label{eq:Cov parsing}
\end{equation}
where $\hat{\lambda}_{ij}=\lambda_{ij}/\sqrt{\pi_i k_{ij}/t}$ is the unit-variance noise source at transition $i\mapsto j$.
Geometrically, this is the inner product projection rule for vectors: $\textbf{u}\cdot \textbf{v} = \sum_{z} (\textbf{u}\cdot\hat{\textbf{e}}_z)(\textbf{v}\cdot\hat{\textbf{e}}_z)$.
Crucially, the RIM relation (Eq.~\ref{eq:RIM}) further operationalizes this geometry. It transforms the covariance projections---which are otherwise difficult to derive from $\textbf{k}$---into simple matrix elements.
Every term in the decomposition becomes explicitly evaluable via the inverse Jacobian $\textbf{A}^{-1}$.\\

\begin{figure}
\begin{centering}
\includegraphics[width=0.95\columnwidth]{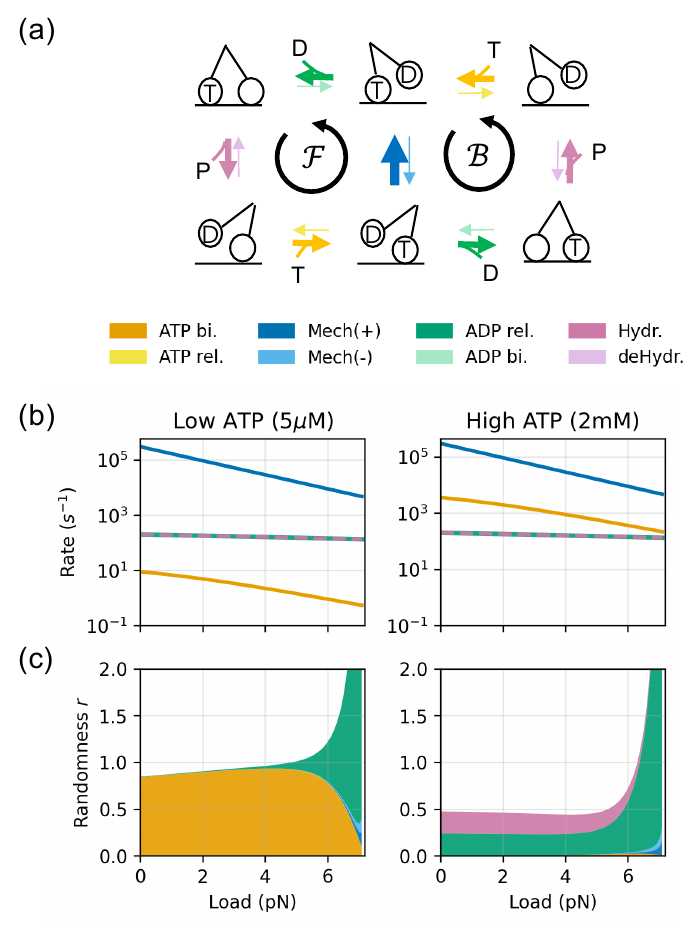}
\par\end{centering}
\caption{\textbf{Dissecting Molecular Motor's Randomness.} \textbf{(a)} The 6-state model for kinesin \citep{liepelt_kinesins_2007}. \textbf{(b)} The bigger the load that the motor has to pull, the slower it runs, for given ATP energy sources \citep{liepelt_kinesins_2007}. \textbf{(c)} Decomposition of the motor's randomness parameter $r$.  The total randomness is decomposed into contributions from functional groups (aggregating forward $\mathcal{F}$ and backward $\mathcal{B}$ cycles), represented by the stacked areas. The upper boundary of the stacked areas are the randomness calculated by the kinesin model, fitted to Visscher et al. \citep{visscher_single_1999}. \label{fig: kinesin}  The motor becomes very inefficient when pulling heavy loads.}
\end{figure}

We now apply Eq.~\eqref{eq:Cov parsing} to a simple model of the kinesin molecular motor to illustrate how the motor ``wastes time'' through structural imprecision.  We define a randomness parameter $r$: it is the variance-to-mean ratio of the net mechanical flux $\psi = (N_+ - N_-)/t$, where $N_\pm$ denotes the forward and backward mechanical step counts.  
We express the motor's randomness in terms of its susceptibility:
\begin{align}
    r=\lim_{t\rightarrow \infty}\frac{t \cdot \mathrm{Var}(\psi)}{\langle \psi \rangle}
    &=\lim_{t\rightarrow \infty} \frac{1}{\langle \psi \rangle}\sum_{i,j}\left(\mathrm{Cov}[\psi,\hat{\lambda}_{ij}]\right)^2 \nonumber \\
    &= \frac{1}{\langle \psi \rangle} \sum_{i,j}  \underbrace{\frac{k_{ij}}{\pi_i}\left( \frac{\partial \langle \psi \rangle}{\partial k_{ij}} \right)^2}_{\text{Sensitivity Contribution}}.
    \label{eq:kinesin_decomp}
\end{align}
This shows that imprecision couples to responsiveness: since the noise contribution scales with the squared sensitivity $(\partial \langle \psi \rangle / \partial k)^2$---a highly responsive step ``amplifies'' its own intrinsic noise into the overall variance. 
Crucially, this result moves beyond bounds like thermodynamic uncertainty relations~\cite{barato_thermodynamic_2015,gingrich_dissipation_2016,horowitz_thermodynamic_2020} which constrain the \textit{magnitude} of flux randomness based on global dynamical time irreversibility. 
Instead, Eq.~\eqref{eq:kinesin_decomp} provides an \textit{exact mechanistic decomposition} in terms of kinetic covariances and sensitivities. 
This approach directly identifies which specific molecular transitions act as the primary sources of the motor's randomness, enabling the targeted dissection of the motor's stochastic mechanism shown below. \\

The power of Eq.~\eqref{eq:kinesin_decomp} is illustrated by a new insight it gives into the molecular bases for the motor's behavior.
Fig.~\ref{fig: kinesin}(b-c) reveals a shift in the stochasticity mechanism of kinesin:
(i) For small loads (far from stalling), the motor is dominated by the \textit{timing noise}, arising from the rate-limiting forward step(s)---the arrival of ATP at low [ATP] or the ADP release and ATP hydrolysis at high [ATP].  (ii) For big loads (near the {stall force}, green region), fluctuations are dominated by a \textit{branching noise}.
That is, near stall, the backward mechanical rate becomes comparable to the forward rate, and the ADP release step dominates the randomness because it controls the critical partitioning between the forward and backward cycles.  This transition in fluctuation properties---between waiting for steps and committing to a step---is not obtainable from approaches that only analyze mean rates; it requires the fluctuation-response relations here.\\

\textbf{The CFT formalism enables scalable design.}
To optimize the randomness parameter $r$ by varying transition rate parameters $k_{ij}$, for example using zero-gradient conditions ($\nabla_k r = 0$) or by driving iterative gradient descent, the computational bottleneck entails evaluating the full gradient of $r$ with respect to all rate constants.  While analytical expressions for this exist, this evaluation becomes computationally prohibitive for large-scale models.  Standard numerical approaches based on generating functions, such as finite differences combined with eigenvalue solvers (e.g., Koza's method \citep{koza_general_1999,barato_thermodynamic_2015}), are costly because they require recomputing the function for every parameter perturbation. \\

Here, we show that our $\textbf{A}^{-1}$ formalism breaks this bottleneck by ensuring algebraic closure: the gradient of the inverse matrix is fully determined by the inverse itself ($\partial \textbf{A}^{-1} = -\textbf{A}^{-1} (\partial \textbf{A}) \textbf{A}^{-1}$).
This allows evaluating both the gradients of mean and variance with a single inverse matrix, significantly improving scalability for locally connected networks---those with $N_{\text{edge}}=\mathcal{O}(N_{\text{node}})$, typical in biophysical models. \\

\begin{figure}
\begin{centering}
\includegraphics[width=1\columnwidth]{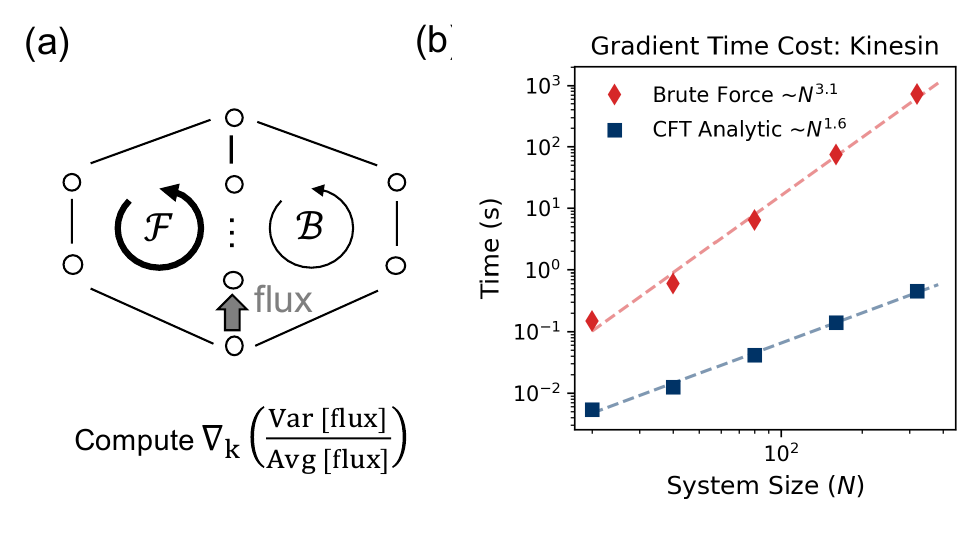}
\par\end{centering}
\caption{\textbf{Breaking the Computational Bottleneck.} \textbf{(a)} The $\Theta$-shape topology used for benchmarking, representing a generalized motor model with multiple mechanical substeps to mimic the diffusive stepping process. \textbf{(b)} Scaling of the computation time for the gradient of the randomness parameter (ratio of variance and average of the flux) as a function of the number of mechanical substeps (and thus the system size). \label{fig: numerical speed up}}
\end{figure}

Again for illustration, we use the kinesin model.  To capture the continuous diffusive nature of the mechanical swing, we extend the single-hop transition in Liepelt and Lipowsky's model \cite{liepelt_kinesins_2007} into a biased random walk across $N$ substeps on a lattice (Fig.~\ref{fig: numerical speed up}a).
We take the chemical rates directly from their original model \cite{SM}.
We benchmarked both methods by computing the full gradient under different numbers of mechanical substeps.
As shown in Fig.~\ref{fig: numerical speed up}(b), the brute-force method (red) scales as $\mathcal{O}(N ^{3.1})$.
In contrast, our $\textbf{A}^{-1}$ approach (blue) scales better, as $\mathcal{O}(N ^{1.6})$, yielding a $>$100-fold speed-up for networks with $N_{\text{node}} \approx 100$.  While our edge-centric algorithm theoretically scales less favorably for dense graphs ($N_{\text{edge}}=\mathcal{O}(N_{\text{node}}^2)$), we show in the Supplemental Material \cite{SM} that it remains computationally superior even on fully connected networks with sizes up to $N_{\text{node}} \approx 100$.\\

\textbf{Universal symmetries and kinetic bounds.}
The geometric structure of the Jacobian $\textbf{A}$ dictates universal principles that govern how any steady-state observables respond to perturbations in any ergodic network flow.  From the identity $\textbf{A}^{-1}\textbf{A} = \textbf{I}$, we derive three sets of local response symmetries, valid for arbitrary CFT observables $\langle x \rangle \in \{\pi_n, \tau_{ij}, J_c\}$. This unifies and generalizes existing response relations \cite{owen_universal_2020,aslyamov_nonequilibrium_2025} within a single principle-based framework. See \cite{SM} for derivations.\\

On each node $n$, we derive a \textbf{Node Escaping Symmetry}:
\begin{equation}
\sum_{l(\neq m)}\frac{k_{ml}}{\pi_{m}}\,\frac{\partial\left\langle x\right\rangle }{\partial k_{ml}}-\sum_{j(\neq n)}\frac{k_{nj}}{\pi_{n}}\,\frac{\partial\left\langle x\right\rangle }{\partial k_{nj}}= \delta_{\langle x \rangle,\pi_n}
\label{eq: dx/dk, from node frequency}
\end{equation}
where the right-hand side is a Kronecker delta, introducing an additional factor of 1 only if $\langle x \rangle = \pi_n$. Physically, this corresponds to perturbing the site energy $E_n$ in an Arrhenius-like rate law ($k_{nj} \propto e^{E_n}$) \cite{owen_universal_2020}, which tunes all outgoing rates simultaneously ($\partial_{E_n} \equiv \sum_{j(\neq n)} k_{nj}\partial_{k_{nj}}$).
Eq. \eqref{eq: dx/dk, from node frequency} thus generalizes the response equalities derived by \citet{owen_universal_2020} from state probabilities to arbitrary CFT observables. Applied to traffic or cycle fluxes, it reveals that these ``node energy'' perturbations scale all network fluxes proportional to local occupancy ($\partial_{E_n} \tau_{ij}  = \pi_n \tau_{ij}$ and $\partial_{E_n} J_c  = \pi_n J_c$).\\

On each edge $ij$, we have an \textbf{Edge Reciprocity}:
\begin{equation}
\frac{1}{\pi_{i}}\,\frac{\partial\left\langle x\right\rangle }{\partial k_{ij}}+\frac{1}{\pi_{j}}\,\frac{\partial\left\langle x\right\rangle }{\partial k_{ji}}= 2 \delta_{\langle x \rangle,\tau_{ij}}.
\label{eq: dx/dk, from traffic}
\end{equation}
Physically, this symmetry governs the barrier height $B_{ij}$ perturbation in an Arrhenius-like rate law  ($k_{ij},k_{ji} \propto e^{-B_{ij}}$) \cite{owen_universal_2020}, which tunes forward and backward rates symmetrically ($\partial_{B_{ij}} \equiv -k_{ij}\partial_{k_{ij}} - k_{ji}\partial_{k_{ji}}$).
Eq. \eqref{eq: dx/dk, from traffic} reveals that the response for $\langle x \rangle \neq \tau_{ij}$ is strictly proportional to the net flux on the perturbed edge $J_{ij}\coloneqq\pi_i k_{ij}-\pi_jk_{ji}$: $\partial_{B_{ij}} \langle x \rangle = -(J_{ij}/\pi_i) \partial_{k_{ij}} \langle x \rangle$.
This implies that barrier perturbations cannot tune distributions $\pi_n$, cycle fluxes $J_c$, or other traffic terms $\tau_{uv}$ ($uv \neq ij)$ if the edge is ``locally'' at equilibrium ($J_{ij}=0$). They acquire control only through a nonzero nonequilibrium net flux $J_{ij}$.
Together with our covariance decomposition in Eq. \eqref{eq:Cov parsing}, these symmetries underpin the flux fluctuation-response identities derived by \citet{aslyamov_nonequilibrium_2025}, while extending their validity to arbitrary CFT observables.\\

On each cycle $c$, we find a \textbf{Cycle Symmetry}:
\begin{equation}
\sum_{ij\in c^{+}}\frac{1}{\pi_{i}}\,\frac{\partial\left\langle x\right\rangle }{\partial k_{ij}}-\sum_{ji\in c^{-}}\frac{1}{\pi_{j}}\,\frac{\partial\left\langle x\right\rangle }{\partial k_{ji}}=2 \delta_{\langle x \rangle,J_c}.
\label{eq: dx/dk, from the cycle}
\end{equation}
This enforces a topological constraint: the cumulative sensitivity along any forward cycle must precisely balance that of its reversal (for observables $\langle x \rangle \neq J_c$, including all $\pi_n,\tau_{ij},$ and the other cycle fluxes $J_{c'}$).\\

\begin{figure}
\begin{centering}
\includegraphics[width=\columnwidth]{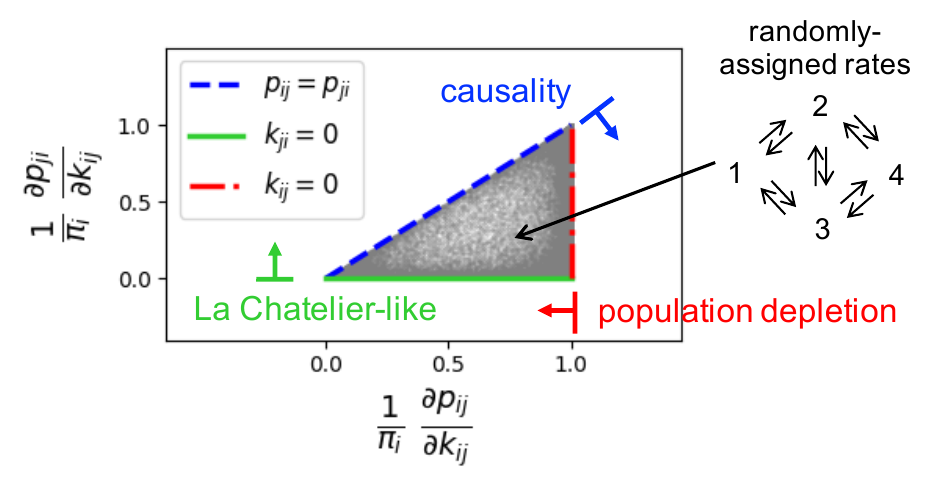}
\par\end{centering}
\caption{\textbf{Universal Kinetic Bounds.} 
The scatter plot numerically validates the derived sensitivity hierarchy using an ensemble of random 4-state networks as an example (schematic, upper right). Each grey dot represents the normalized response of the forward flux $p_{ij}$ (x-axis) versus the induced response of the reverse flux $p_{ji}$ (y-axis) to a perturbation in the forward rate $k_{ij}$. The feasible response region is strictly bounded by three physical limits: {Population Depletion}, {Causality}, and {Le Chatelier-like Compensation}. \label{fig: RR}}
\end{figure}

Our framework also provides the physical origins of established response bounds.
By identifying the algebraic coefficients $\Delta_{ij}$ and $\nabla_{ij}$ defined in \citet{aslyamov_nonequilibrium_2024} as specific rate sensitivities \cite{SM}:
\begin{align}
\pi_{i}(1-\Delta_{ij})=\frac{\partial J_{ij}}{\partial k_{ij}} \quad \text{and} \quad \pi_{i}(1+\nabla_{ij})=\frac{\partial \tau_{ij}}{\partial k_{ij}},\label{eq: Delta and Nabla}
\end{align}
we elevate their algebraic constraints ($0\le\Delta_{ij}\le1$ and $|\nabla_{ij}|\le\Delta_{ij}$) into a physical kinetic hierarchy for the one-way fluxes $p_{ij} = \pi_i k_{ij}$:
\begin{equation}
\pi_{i}\ge\frac{\partial p_{ij}}{\partial k_{ij}}\ge\frac{\partial p_{ji}}{\partial k_{ij}}\ge 0.
\label{eq: pij inequalities}
\end{equation}
As visualized in the triangular domain of Fig.~\ref{fig: RR}, this hierarchy reflects the physical mechanisms governing nonequilibrium flow response, valid for any steady-state network flows.\\

The upper bound, $\pi_i \ge \partial_{k_{ij}} p_{ij} = \pi_i + k_{ij} ~\partial_{k_{ij}} \pi_{i}$, reflects \textit{population depletion:} since increasing the escape rate $k_{ij}$ drains the source population ($\partial_{k_{ij}}\pi_i \le 0$), the flux cannot grow faster than the population size $\pi_i$. This limit is saturated as $k_{ij} \to 0$, where the perturbation is too weak to shift the steady-state population.
The middle inequality, $\partial_{k_{ij}} p_{ij} \ge \partial_{k_{ij}} p_{ji}$, reflects \textit{causality} ($\partial_{k_{ij}} J_{ij} \ge 0$), showing a stronger push always increases net flux in the driven direction. This bound tightens to equality ($p_{ij}=p_{ji}$) at zero-flux edges, characteristic of detailed balance or a topological bridge.
Finally, the non-negativity of the induced reversed response ($\partial_{k_{ij}} p_{ji} \ge 0$) captures a \textit{Le Chatelier-like compensation}: pushing ``particles'' into the target state $j$ increases its occupancy, which in turn drives an increased reverse backflow even when the reverse rate $k_{ji}$ remains fixed.\\

\textbf{Conclusions}: We have presented a force-based framework for nonequilibrium fluctuations and responses.  It bridges between formal variational theory and computational utility. These relations reflect a deeper observable-force conjugacy, encoded in the inverse of a single Jacobian matrix. We demonstrated its dual utility: Analytically, it reveals principles of design and optimization, as shown in the kinesin molecular motor.  Computationally, it scales efficiently for large networks. And it reveals general response relations and universal constraints on stochastic flows.\\

\textbf{Code Availability:} The codes used to produce the Figures
are available at https://doi.org/10.5281/zenodo.18611573.\\

\begin{acknowledgments}
\textbf{Acknowledgments: }We are grateful for the financial support from
the Laufer Center for Physical and Quantitative Biology at Stony Brook,
the John Templeton Foundation (Grant ID 62564), and NIH (Grant RM1-GM135136).
\end{acknowledgments}

\bibliography{Paper-References}

\clearpage
\onecolumngrid 

\setcounter{equation}{0}
\setcounter{figure}{0}
\setcounter{table}{0}
\setcounter{page}{1}
\setcounter{section}{0}

\renewcommand{\theequation}{S\arabic{equation}}
\renewcommand{\thefigure}{S\arabic{figure}}
\renewcommand{\thetable}{S\arabic{table}}

\begin{center}
    {\large \textbf{Supplemental Material for:\\ ``Fluctuation-Response Design Rules for Nonequilibrium Flows''}} \\
    
    \vspace{1em} 
    
    Ying-Jen Yang$^{1,*}$ and Ken A. Dill$^{1,2,3}$ \\
    
    \vspace{0.5em} 
    
    {\small \itshape
    $^{1}$Laufer Center of Physical and Quantitative Biology, Stony Brook University \\
    $^{2}$Department of Physics, Stony Brook University \\
    $^{3}$Department of Chemistry, Stony Brook University} \\
    
    \vspace{0.5em}
    
    {\small $^*$ying-jen.yang@stonybrook.edu} \\

    \vspace{1em} 

    (Dated: \today)
\end{center}

\tableofcontents{}

\title{Supplementary Material for ``Fluctuation-Response Duality for Scalable Nonequilibrium Design''}
\author{Ying-Jen Yang}
\email{ying-jen.yang@stonybrook.edu}

\affiliation{Laufer Center of Physical and Quantitative Biology, Stony Brook University}
\author{Ken A. Dill}
\affiliation{Laufer Center of Physical and Quantitative Biology, Stony Brook University}
\affiliation{Department of Physics, Stony Brook University}
\affiliation{Department of Chemistry, Stony Brook University}

\section{The Caliber Force Framework and the Sparse Jacobian $\mathbf{A}$}

\subsection{Stochastic Path Entropy of Markov Jump Processes}

The path probability of Markov Jump processes with transition rates $k_{ij}$ (from $i$ to $j$) has an explicit expression. For path $\omega_t$ with time length $t$, it is \begin{equation}
    \mathcal{P}_{\mathbf{k}} (\omega_t)=p_0 (i_0) [e^{k_{i_0i_0} (t_1-0)} k_{i_0 i_1}]\cdots [e^{k_{i_{n-1} i_{n-1}} (t_n-t_{n-1})} k_{i_{n-1} i_n}]e^{k_{i_n i_n} (t-t_{n})}
\end{equation}
where $p_0$ is the initial distribution, $t_n$ is the $n$-th jump times, and $k_{ii} = -\sum_{j(\neq i)} k_{ij}$ is the negative of the escape rate at state $i$. By using counting variables---$N_{ij}(\omega_t)$ as the number of $i\mapsto j$ jumps and $T_i$ as the total dwell time in state $i$---the path probability has a simpler form: 
\begin{equation}
    \mathcal{P}_{\mathbf{k}} (\omega_t)=p_0 (i_0) \exp\left(\sum_{i\neq j}N_{ij}[\omega_{t}]\ln k_{ij} -\sum_{i}T_{i}[\omega_{t}]\sum_{j(\neq i)}k_{ij}\right).
\end{equation}
The \textit{stochastic relative entropy} between the process with rates $k_{ij}$ and a reference process with unit rates $u_{ij}=1$ (both with the same initial distribution) is then
\begin{equation}
    \mathcal{R}[\omega_t;\mathbf{k}]=\ln\frac{\mathcal{P}_{\mathbf{k}}(\omega_{t})}{\mathcal{P}_{\mathbf{u}}(\omega_{t})}=\sum_{i\neq j}N_{ij}(\omega_{t})\ln k_{ij}-\sum_{i}T_{i}(\omega_{t})~\varepsilon_i(\mathbf{k}).
    \label{eq: R in terms of N and T}
\end{equation}
where $\varepsilon_i = \sum_{j(\neq i)}k_{ij}-n_i$ is the escape rate offset by the out-degree of state $i$.
This expression holds for arbitrary $t$. In mathematics, the path probability ratio (exponential of stochastic entropy) is also known as the \textit{Radon-Nikodym derivatives}, typically denoted as $\text{d}\mathbb{P}_{\mathbf{k}}/\text{d}\mathbb{P}_{\mathbf{u}}$ in measure-theoretic probability theory. It captures all statistical differences between two processes ($\mathbf{k}$ and $\mathbf{u}$) as the expected value of any path observables $f(\omega_t)$ can be connected by \begin{equation}
    \langle f \rangle_{\mathbf{k}} = \sum_{\omega_t} \mathcal{P}_\mathbf{k} (\omega_t) f(\omega_t) = \sum_{\omega_t} \mathcal{P}_\mathbf{u} (\omega_t) \cdot \frac{\mathcal{P}_\mathbf{k} (\omega_t)}{\mathcal{P}_\mathbf{u} (\omega_t)} f(\omega_t) =\langle e^\mathcal{R} f\rangle_\mathbf{u}.
\end{equation}
Identifying time irreversibility (entropy production) as a type of stochastic relative entropy (with specific choices of reversals as the reference process) leads to their various fluctuation theorems (for a recent review, see \cite{yang_unified_2020}).

\subsection{Fundamental Observables and the Asymptotic Constraints from Conservation}
The counting variables $(N_{ij},T_i)$ are elemental as any time-extensive first-order path observables can be expressed by them: \begin{align}
    \mathcal{B}(\omega_t) = \int_0^t~f(i_t) ~\text{d}t + \sum_{n} g(i_{t_{n-1}}, i_{t_n}) = \sum_i T_i~ f(i) + \sum_{i \neq j}N_{ij}~ g(i,j)
\end{align} where $i_t$ is the state at time $t$ and $n$ sums over all jumps. However, they are degenerate due to the conservation of probability. We always have normalization constraints, $\sum_{i} T_i=t$, and in, the long-term, the transition counts must satisfy inward/outward flux balance (Kirchhoff's current law)  \cite{barato_formal_2015}: $\sum_{j(\neq i)} N_{ij}= \sum_{j(\neq i)} N_{ji} + o(t)$ where $o(t)$ represents sublinear term $\lim_{t\rightarrow \infty} o(t)/t =0$, which can also be denoted as $\sum_{j(\neq i)} N_{ij}\sim \sum_{j(\neq i)} N_{ji}$ by using standard asymptotic analysis notation $a\sim b \Leftrightarrow \lim_{t\rightarrow \infty} a/b=1$.\\

The flux balance indicates that net fluxes $N_{ij}-N_{ji}$ are cyclic in the long term. Thus, to handle the constraints, we follow Caliber Force Theory (CFT) \cite{yang_principled_2025} and use the counting variables $\textbf{X}=(\Phi_{ij},T_n,\Psi_c)$. Traffic $\Phi_{ij}=N_{ij}+N_{ji}$ is the total flux, orthogonal to the net fluxes; the dwell time $T_n$ excludes an arbitrary chosen reference node $m$ in the network to handle normalization; $\Psi_c=N_{ab}-N_{ba}$ is the fundamental cycle flux where $ab$ is the defining (directed) chord of the fundamental cycle $c$ (obtained by adding one chord to an arbitrarily-chosen spanning tree of the state space network). The net flux on a given edge is simply the sum of the fundamental cycle fluxes that circulate through the edge: \begin{equation}
    N_{ij} - N_{ji} \sim \sum_c \Theta_{ij}^c \Psi_c
\end{equation}
where $\Theta_{ij}^c$ is the incidence matrix---it equals to 1 if edge $ij$ aligns with the directed cycle $c$, -1 if the reverse $ji$ aligns with the cycle $c$, and 0 if $ij$ doesn't belong to cycle $c$. 
The set of fundamental counting observables $\textbf{X}=(\Phi_{ij},T_n,\Psi_c)$ thus spans $(N_{ij},T_i)$. This exploits the identity $N_{ij} = \frac{1}{2}[(N_{ij}+N_{ji}) + (N_{ij}-N_{ji})]$, which decomposes any count into a symmetric traffic part and an antisymmetric net flux part:
\begin{equation}
    N_{ij} \sim \frac{\Phi_{ij}+\sum_c \Theta_{ij}^c \Psi_c}{2},~T_i = \begin{cases}
T_n & \text{, if }i\neq m\\
t-\sum_{n(\neq m)} T_n & \text{, if }i= m
\end{cases}.\label{eq: spanning N,T with X}
\end{equation} and thus all time extensive observables $\mathcal{B}(\omega_t)$.

\subsection{Conjugated Forces from Caliber Force Theory}
The stochastic path entropy $\mathcal{R}[\omega_t;\mathbf{k}]$ is an example of time-extensive observable $\mathcal{B}(\omega_t)$. We can thus rewrite it in terms of the fundamental counting variables $\textbf{X}=(\Phi_{ij},T_n,\Psi_c)$. Plugging Eq. \eqref{eq: spanning N,T with X} into Eq. \eqref{eq: R in terms of N and T} leads to 
\begin{align}
    &\mathcal{R}[\omega_t;\mathbf{k}]
     \sim \sum_{i\neq j}\frac{\Phi_{ij}+\sum_c\Theta_{ij}^c \Psi_c}{2}\ln k_{ij}-\sum_{n}T_{n}(\omega_{t}) ~\varepsilon_n-T_m(\omega_{t}) ~\varepsilon_m\\
    &= \sum_{i\neq j}\frac{\Phi_{ij}}{2}\ln k_{ij}+\sum_{i\neq j}\frac{\sum_c\Theta_{ij}^c \Psi_c}{2}\ln k_{ij}-\sum_{n}T_{n}(\omega_{t}) ~\varepsilon_n-\left(t-\sum_{n(\neq m)} T_n(\omega_{t}) \right) ~\varepsilon_m. 
\end{align}
Using the symmetry of $\Phi_{ij}=\Phi_{ji}$ and exchanging the two summation of the second term, we get the asymptotic work-like form \begin{align}
    \mathcal{R}[\omega_t;\mathbf{k}] &\sim \sum_{i<j} \Phi_{ij}~ \mathfrak{F}_{\text{edge},ij}+\sum_c \Psi_c ~\mathfrak{F}_{\text{cycle},c}+\sum_n T_n ~\mathfrak{F}_{\text{node},n} - t~ \mathfrak{c}\\
    &=\textbf{X}(\omega_t) \cdot \boldsymbol{\mathfrak{F}}(\mathbf{k})- t~\mathfrak{c}(\mathbf{k}).\label{eq: R as forces and caliber}
\end{align}
Here, the forces $\boldsymbol{\mathfrak{F}}(\mathbf{k})$  are 
\begin{align}
\mathfrak{F}_{\text{edge},ij} & =\frac{1}{2}\ln{k_{ij}k_{ji}}, \\
\mathfrak{F}_{\text{node},n} & =\sum_{i(\neq m)}(k_{mi}-1)-\sum_{j(\neq n)}(k_{nj}-1), \\
\mathfrak{F}_{\text{cycle},c} & =\frac{1}{2}\ln{\frac{k_{i_{0}i_{1}}k_{i_{1}i_{2}}\cdots k_{i_{\sigma}i_{0}}}{k_{i_{0}i_{\sigma}}k_{i_{\sigma}i_{\sigma-1}}\cdots k_{i_{1}i_{0}}}}\label{eqs: force-expression-in-terms-of-the-rates}
\end{align} where $i_0i_1...i_\sigma i_0$ is the state sequence of the cycle $c$.
The caliber (rate) $\mathfrak{c}(\mathbf{k})=\varepsilon_m = \sum_{j(\neq m)}k_{im} - n_m$ has the value of the escape rate of the reference node $m$.\\

Eq. \eqref{eq: R as forces and caliber} can be understood as follows. The stochastic path entropy $\mathcal{R}$ on the left quantifies the difference between the current process with rate $\mathbf{k}$ and a reference process with unit rate. The forces on the right are affinities driving edge exchange, node escaping, and cycle completion, with the counting variables $\textbf{x}$ giving the statistical occurrence weights on the forces. The caliber term $(\mathfrak{c}~ t)$ is a ``renormalization'' factor. In fact, it has be shown in CFT \cite{yang_principled_2025} that the caliber is the log partition function for dynamics: \begin{align}
    \lim_{t\rightarrow\infty} \frac{1}{t} \ln \sum_{\omega_t} \mathcal{P}_{\mathbf{u}}(\omega_t) e^{\boldsymbol{\mathfrak{F}}\cdot\textbf{X}(\omega_t)} &=\lim_{t\rightarrow\infty} \frac{1}{t} \ln \sum_{\omega_t} \mathcal{P}_{\mathbf{k}}(\omega_t) e^{-\mathcal{R}} e^{\boldsymbol{\mathfrak{F}}\cdot\textbf{X}(\omega_t)} \nonumber\\ 
    &= \lim_{t\rightarrow\infty} \frac{1}{t} \ln e^{t \mathfrak{c}}\sum_{\omega_t} \mathcal{P}_{\mathbf{k}}(\omega_t) \nonumber\\
    &= \mathfrak{c}(\boldsymbol{\mathfrak{F}}).\label{eq: caliber as log partition function}
\end{align}
Furthermore, its Legendre transform gives the average path entropy rate \cite{yang_principled_2025}: \begin{equation}
    -\mathfrak{s}_{\text{path}} = \lim_{t\rightarrow \infty} \frac{\mathcal{R}}{t} = \boldsymbol{\mathfrak{F}}\cdot \langle \textbf{x} \rangle -\mathfrak{c} \label{eq: Legendre transform}
\end{equation}
where $\textbf{x} = \textbf{X}/t$ is the vector of the observable rates and $\langle \textbf{x} \rangle$ is the vector of their steady-state averages, and the forces emerge naturally as the thermodynamics-like conjugate variables: \begin{equation}
    \boldsymbol{\mathfrak{F}} = -\frac{\partial \mathfrak{s}_{\text{path}}}{\partial \langle \textbf{x} \rangle}.
\end{equation} Eq. \eqref{eq: R as forces and caliber} shown in this paper is, in fact, the stochastic version of Eq. \eqref{eq: Legendre transform} derived in \cite{yang_principled_2025}, based on the law of large number in the long term limit $t\rightarrow \infty$: \begin{equation}
    \frac{\mathcal{R}}{t}\rightarrow (-\mathfrak{s}_{\text{path}})\text{ and } \frac{\textbf{X}}{t}\rightarrow \langle \textbf{x}\rangle.
\end{equation}

\subsection{Generic Construction of the Jacobian matrix $\mathbf{A}$}
We now consider how the stochastic path entropy $\mathcal{R}$ (Eqs. \ref{eq: R in terms of N and T} and \ref{eq: R as forces and caliber}) varies when changing the process $\mathbf{k}$ of interest, under a fixed $\omega_t$ (and $t$):
\begin{align}
\frac{\partial \mathcal{R}}{\partial \ln k_{ij}} &= (N_{ij} - T_i k_{ij}) \label{eq: dR/dlnk = t lambda}\\
&\sim\textbf{x}\cdot \frac{\partial \boldsymbol{\mathfrak{F}}}{\partial \ln k_{ij}} - t\frac{\partial \mathfrak{c}}{\partial \ln k_{ij}}.
\end{align}
Thus, denoting $\lambda_{ij} = (N_{ij}-T_i k_{ij})/t$ and $\textbf{x}=\textbf{X}/t$, we arrive the key linear relationship between these Poissonian fluctuation on each transition $ij$ and the CFT observable (rate) $\textbf{x}:$ \begin{equation}
    \boldsymbol{\lambda} (\omega_t) \sim \mathbf{A}(\textbf{k})~ \textbf{x}(\omega_t) -\nabla_{\ln \mathbf{k}} \mathfrak{c}(\textbf{k})\label{eq: affine relation}
\end{equation}
where $\mathbf{A}_{(ij),\beta} = \partial \mathfrak{F_\beta}/\partial \ln k_{ij}.$
One could also derive this linear map directly from how $\textbf{X}=(\Phi_{ij},T_n,\Psi_c)$ spans $(N_{ij},T_i)$ with Eq. \eqref{eq: spanning N,T with X}, but that derivation obscures the physical origin of $\mathbf{A}$'s sparsity. In our framework, the sparsity of $\mathbf{A}$ arises explicitly from the locality of the CFT forces $\boldsymbol{\mathfrak{F}}$ (Eqs. \ref{eqs: force-expression-in-terms-of-the-rates}), where each force depends only on a small, local subset of transition rates.\\

By defining $\lambda'_{mj} = \lambda_{mj} + (\partial \mathfrak{c}/\partial \ln k_{mj}) = \lambda_{mj}+k_{mj}$ and denoting $\textbf{x} = \textbf{X}/t=(\Phi_{ij}/t,T_n/t,\Psi_c/t) = (\phi_{ij},f_n,\psi_c)$, we can rewrite Eq. \eqref{eq: affine relation} into a simpler linear map:  \begin{equation}
    \tilde{\boldsymbol{\lambda}} \sim \mathbf{A} \textbf{x}
\end{equation}
where $\tilde{\boldsymbol{{\lambda}}}$ has components $\lambda_{ij}$ for $i(\neq m)$ and $\lambda'_{ij}$ for $i=m$. By ordering the edge properly, the sparse matrix $\mathbf{A}$ has a generic form as shown in Fig. \ref{fig:general A construction} (an example is given in the main text):
\begin{figure}[h!]
    \centering
    \includegraphics[width=0.7\linewidth]{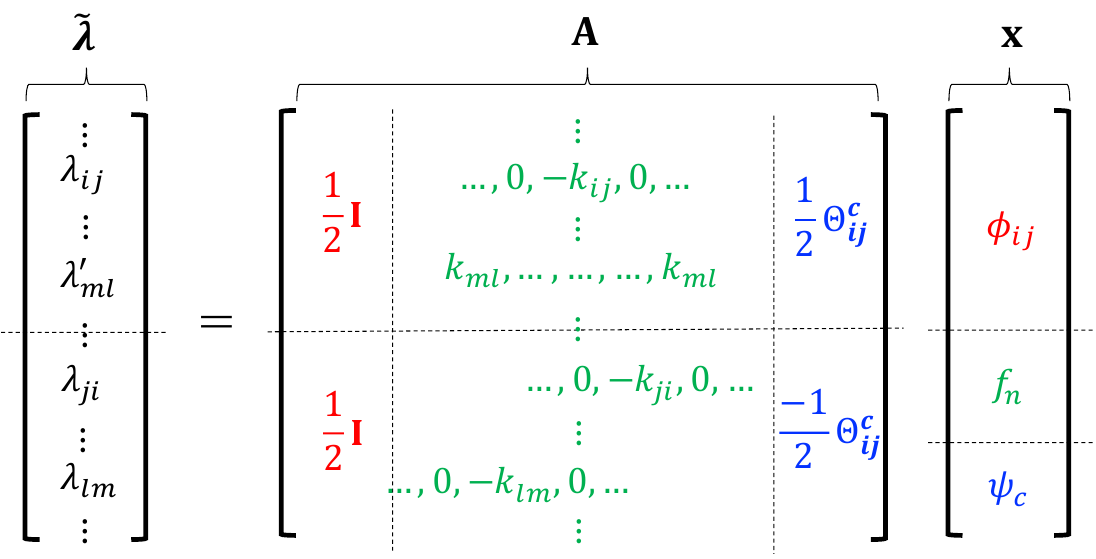}
    \caption{The matrix maps the CFT observable vector $\textbf{x} = ({\phi}, f, {\psi})^{\mathsf{T}}$ to the vector of independent noise sources $\tilde{\boldsymbol{\lambda}}$. The sparse geometry consists of three distinct vertical blocks: (Left, Red) a diagonal identity block scaling edge traffic; (Middle, Green) a sparse block containing transition rates linking node frequencies to noise; and (Right, Blue) a topological block containing the cycle incidence matrix $\Theta_{ij}^c$. The vector $\tilde{\boldsymbol{\lambda}}$ accounts for the shifted noise $\lambda'_{ml}$ for transitions originating from the reference state $m$.}
    \label{fig:general A construction}
\end{figure}

\section{The Response-Inverse-Matrix (RIM) relation}

We here derive one of our main result: \begin{equation}
    \frac{\partial \langle x_\alpha \rangle}{\partial \ln k_{ij}} = t~\mathrm{Cov}[x_\alpha,\lambda_{ij}] \sim \pi_i k_{ij}~ [\mathbf{A}^{-1}]_{\alpha,(ij)} = \pi_i k_{ij}~ \nabla_{\mathfrak{F}_\alpha}\ln {k_{ij}}.
\end{equation}
The first equality is known and recently discussed by Zheng and Lu \cite{zheng_unified_2025}. We show here that, for CFT observables $\textbf{x}$, all these responses and covariances are encoded in the inverse of a sparse Jacobian $\mathbf{A}$.

\subsection{The General Score Function Derivation}
Consider a generic path observable $f(\omega_t)$ whose functional form is independent to $k_{ij}$. It is straightforward to show that the average susceptibility to $\ln k_{ij}$ perturbation (i.e. $\partial_{\ln k}\equiv k\partial_k$) is the covariance between $f$ and the independent noise sources $\lambda_{ij}$: \begin{align}
   & k_{ij} \frac{\partial}{\partial k_{ij}} \sum_{\omega_t} \mathcal{P}_{\mathbf{k}}(\omega_t) f(\omega_t) = \sum_{\omega_t} \frac{\partial \mathcal{P}_{\mathbf{k}}}{\partial \ln k_{ij}}~ f(\omega_t) = \sum_{\omega_t} \mathcal{P}_\mathbf{k} (\omega_t)~\frac{\partial \ln \mathcal{P}_{\mathbf{k}}}{\partial \ln k_{ij}}~ f(\omega_t) \\
    &= \sum_{\omega_t} \mathcal{P}_\mathbf{k} (\omega_t)~\frac{\partial \mathcal{R}}{\partial \ln k_{ij}}~ f(\omega_t) =t\sum_{\omega_t} \mathcal{P}_\mathbf{k} (\omega_t)~\lambda_{ij}~ f(\omega_t) = t~ \mathrm{Cov} (\lambda_{ij},f)
\end{align}
where we have used Eq. \eqref{eq: dR/dlnk = t lambda} and that $\lambda_{ij}$ is zero mean.
Applying it to CFT observables $\textbf{x}$ and taking the long-term limit $t\rightarrow \infty$ leads to the relation between steady-state average responses and asymptotic covariances: \begin{equation}
    \frac{\partial \langle x_\alpha\rangle}{\partial \ln k_{ij}} = \lim_{t\rightarrow \infty}t~ \mathrm{Cov} (\lambda_{ij},x_\alpha) \label{eq: fluc-resp to x}
\end{equation}
where $x_\alpha$ can be any of $(\phi_{ij},f_n,\psi_c).$\\

To link these to the Jacobian $\mathbf{A}=\nabla_{\ln \mathbf{k}}\boldsymbol{\mathfrak{F}}$, we use the (asymptotic) linear relation \begin{equation}
    \boldsymbol{\lambda} - \nabla_{\ln \mathbf{k}} \mathfrak{c} \sim \mathbf{A}\textbf{x} \Rightarrow\textbf{x}\sim \mathbf{A}^{-1} (\boldsymbol{\lambda} - \nabla_{\ln \mathbf{k}} \mathfrak{c}). \label{eq: inv A to get x}
\end{equation}
 The inverse matrix $\mathbf{A}^{-1}$ exists because CFT forces constitute a coordinate for the parameter space \cite{yang_principled_2025}, i.e. $\mathbf{k}\leftrightarrow \boldsymbol{\mathfrak{F}}$ is one-to-one. Plugging Eq. \eqref{eq: inv A to get x} into Eq. \eqref{eq: fluc-resp to x} leads to 
\begin{equation}
    \frac{\partial \langle x_\alpha\rangle}{\partial \ln k_{ij}} = \lim_{t\rightarrow \infty}t~ \mathrm{Cov} (\lambda_{ij},[\mathbf{A}^{-1}(\boldsymbol{\lambda}-\nabla_{\ln \mathbf{k}}\mathfrak{c})]_{\alpha}). \label{eq: fluc-resp to x lim t}
\end{equation}
Now, we evaluate the covariance of the noise sources $\lambda_{ij} = (N_{ij}-T_i k_{ij})/t$. Since elementary jumps follow Poisson statistics in the short-time limit and remain independent across different edges, the variance of the count $N_{ij}$ scales with its mean $t \pi_i k_{ij}$. This yields the diagonal covariance structure \cite{zheng_unified_2025}: \begin{equation}
    \lim_{t\to\infty} t \cdot \mathrm{Cov}[\lambda_{ij},\lambda_{ab}] = \langle N_{ij}/t \rangle \delta_{i,a} \delta_{j,b} = \pi_i k_{ij} ~ \delta_{i,a} ~ \delta_{j,b}.
\end{equation}
This reduces Eq. \eqref{eq: fluc-resp to x lim t} into 
\begin{equation}
    \frac{\partial \langle x_\alpha\rangle}{\partial \ln k_{ij}} = \pi_i k_{ij} [\mathbf{A}^{-1}]_{\alpha,(ij)}. \label{eq: RIM}
\end{equation}

\subsection{The Alternative Derivation from Caliber Force Theory (CFT)}
Here, we provide an alternative derivation from CFT, demonstrating the consistency of RIM with CFT's fluctuation-response equalities \cite{yang_principled_2025}.
Given that the caliber $\mathfrak{c}(\boldsymbol{\mathfrak{F}})$ functions as the log partition function (as revisited in Eq. \ref{eq: caliber as log partition function}), its first derivatives yield steady-state averages:
\begin{equation}
    \langle x_\alpha \rangle = \frac{\partial \mathfrak{c}}{\partial \mathfrak{F}_\alpha},
\end{equation}
and its second derivatives yield covariances:
\begin{align}
\frac{\partial\langle x_{\alpha}\rangle}{\partial\mathfrak{F}_\beta} =\frac{\partial\langle x_{\beta}\rangle}{\partial\mathfrak{F}_\alpha}= \frac{\partial^2 \mathfrak{c}}{\partial \mathfrak{F}_\alpha \partial \mathfrak{F}_\beta}=\lim_{t\rightarrow\infty} & t\cdot{\rm Cov}(x_\alpha,x_\beta)\label{eq: CFT FRR}
\end{align}
where $(x_\alpha,x_\beta)$ can be any pair from $(f_{n},\phi_{ij},\psi_{c})$, and $\mathfrak{F}_\alpha$ is the conjugate force of $x_\alpha$.\\

Consider a rate perturbation of an arbitrary mean observable, $\partial\langle x\rangle/\partial k_{ij}$.
When $i\neq m$, this perturbation affects the node force $\mathfrak{F}_{i}$, the edge force $\mathfrak{F}_{ij}$, and the cycle forces $\mathfrak{F}_{c}$ for all cycles $c$ passing through edge $ij$. Therefore, in the force coordinate system, we can expand the rate responses in terms of CFT force responses:
\begin{align}
\frac{\partial\langle x\rangle}{\partial k_{ij}}= & \frac{\partial\langle x\rangle}{\partial\mathfrak{F}_{i}}\frac{\partial\mathfrak{F}_{i}}{\partial k_{ij}}+\frac{\partial\langle x\rangle}{\partial\mathfrak{F}_{ij}}\frac{\partial\mathfrak{F}_{ij}}{\partial k_{ij}}+\sum_{c}\frac{\partial\langle x\rangle}{\partial\mathfrak{F}_{c}}\frac{\partial\mathfrak{F}_{c}}{\partial k_{ij}}\nonumber \\
= & \frac{\partial\langle x\rangle}{\partial\mathfrak{F}_{i}}(-1)+\frac{\partial\langle x\rangle}{\partial\mathfrak{F}_{ij}}\left(\frac{1}{2k_{ij}}\right)+\sum_{c}\frac{\partial\langle x\rangle}{\partial\mathfrak{F}_{c}}\left(\frac{1}{2k_{ij}}\Theta_{c}^{ij}\right)\label{eq: dx/dk in forces step 1}
\end{align}
where $\Theta_{c}^{ij}$ is the (directed) incidence matrix: it equals $1$ if $ij$ aligns with $c$, $-1$ if $ij$ anti-aligns with $c$, and $0$ otherwise.
Now, using the fluctuation-response relations in CFT:
\begin{align}
\frac{\partial\langle x\rangle}{\partial\mathfrak{F}_{i}}=\lim_{t\rightarrow\infty} & t\cdot{\rm Cov}(x,f_{i})\label{eq: node force}\\
\frac{\partial\langle x\rangle}{\partial\mathfrak{F}_{ij}}=\lim_{t\rightarrow\infty} & t\cdot{\rm Cov}(x,f_{ij}+f_{ji})\label{eq: edge force}\\
\frac{\partial\langle x\rangle}{\partial\mathfrak{F}_{c}}=\lim_{t\rightarrow\infty} & t\cdot{\rm Cov}(x,f_{ab}-f_{ba}).\label{eq: cycle force}
\end{align}
However, since asymptotic flux balance implies $\lim_{t\rightarrow\infty}\sum_{j}(f_{ij}-f_{ji})=0$, the net edge flux $f_{ij}-f_{ji}$ is asymptotically spanned by the fundamental cycle fluxes:
\begin{align}
\lim_{t\rightarrow\infty}(f_{ij}-f_{ji}) & -\sum_{c}\Theta_{c}^{ij}(f_{ab}-f_{ba})=0.\label{eq: asymptotic spanning of edge net flux with cycles}
\end{align}
Substituting these into Eq. \eqref{eq: dx/dk in forces step 1}, we obtain:
\begin{align*}
\frac{\partial\langle x\rangle}{\partial k_{ij}}= & \lim_{t\rightarrow\infty}t\cdot\left[-{\rm Cov}(x,f_{i})+\frac{1}{2k_{ij}}\left({\rm Cov}(x,f_{ij}+f_{ji})+\sum_{c}\Theta_{c}^{ij}{\rm Cov}(x,f_{ab}-f_{ba})\right)\right]\\
= & \lim_{t\rightarrow\infty}t\cdot\left[-{\rm Cov}(x,f_{i})+\frac{1}{2k_{ij}}\left({\rm Cov}(x,f_{ij}+f_{ji})+{\rm Cov}(x,f_{ij}-f_{ji})\right)\right]\\
= & \lim_{t\rightarrow\infty}t\cdot\left[-{\rm Cov}(x,f_{i})+\frac{1}{k_{ij}}{\rm Cov}(x,f_{ij})\right].
\end{align*}
Thus, by identifying $\lambda_{ij}=f_{ij}-f_{i}k_{ij}$, we have shown (for $i\neq m$):
\begin{align*}
k_{ij}\frac{\partial\langle x\rangle}{\partial k_{ij}} & =\lim_{t\rightarrow\infty}t\cdot{\rm Cov}[x,\lambda_{ij}].
\end{align*}
When $i=m$, the perturbation $k_{ij}$ affects all node forces (while the edge and cycle contributions remain identical to the $i \neq m$ case, denoted by ``$\dots$''). We get:
\begin{align}
\frac{\partial\langle x\rangle}{\partial k_{ij}} & =\sum_{n(\neq i)}\frac{\partial\langle x\rangle}{\partial\mathfrak{F}_{n}}\underbrace{\frac{\partial\mathfrak{F}_{n}}{\partial k_{ij}}}_{=1}+\cdots\nonumber \\
 & =\lim_{t\rightarrow\infty}t\sum_{n(\neq i)}{\rm Cov}(x,f_{n})+\cdots\nonumber \\
 & =\lim_{t\rightarrow\infty}t\cdot{\rm Cov}\left(x,\sum_{n(\neq i)}f_{n}\right)+\cdots\nonumber \\
 & =\lim_{t\rightarrow\infty}t\cdot{\rm Cov}(x,1-f_{i})+\cdots\nonumber \\
 & =-\lim_{t\rightarrow\infty}t\cdot{\rm Cov}(x,f_{i})+\cdots\label{eq: i=m case}
\end{align}
where we have used the linearity of covariance and the fact that a constant shift in the variable does not change the covariance. Eq. \eqref{eq: i=m case} completes the CFT-based derivation for Eq. (12) of the main text.

\subsection{Derivation of the Covariance Decomposition}
First, we note that the covariance is invariant under constant shifts, i.e., ${\rm CoV}[X, Y+c] = {\rm CoV}[X, Y]$. This implies that the shifted noise sources at the reference node, $\lambda'_{mj} = \lambda_{mj} + k_{mj}$, contribute identically to the covariance as the raw noise $\lambda_{mj}$. We can thus treat all transitions uniformly in the covariance calculation.\\

The covariance expansion in terms of edge responses can be derived with a direct calculation using the RIM relation: 
\begin{align*}
\lim_{t\rightarrow\infty}t\cdot{\rm CoV}[x,x'] & = \lim_{t\rightarrow\infty}t\cdot{\rm CoV}[\sum_{ab}A_{x,ab}^{-1}\lambda_{ab},\sum_{ij}A_{x',ij}^{-1}\lambda_{ij}]\\
 & = \sum_{ab}\sum_{ij}\lim_{t\rightarrow\infty}t\cdot A_{x,ab}^{-1} A_{x',ij}^{-1} {\rm CoV}[\lambda_{ab},\lambda_{ij}]\\
 & = \sum_{ab}\sum_{ij} A_{x,ab}^{-1}A_{x',ij}^{-1} \left( \pi_{i}k_{ij}\delta_{ai}\delta_{bj} \right) \\
 & = \sum_{ij}\pi_{i}k_{ij}A_{x,ij}^{-1}A_{x',ij}^{-1}.
\end{align*}\\
This gives fast evaluation of covariances with $\mathbf{A}^{-1}.$\\

To explicitly recover the geometric interpretation (Result 1 in the main text), we introduce the \textbf{normalized noise source} $\hat{\lambda}_{ij}$, defined by rescaling $\lambda_{ij}$ with its asymptotic standard deviation $\sigma_{ij} = \sqrt{\pi_i k_{ij}/t}$:
\begin{equation}
    \hat{\lambda}_{ij} \equiv \frac{\lambda_{ij}}{\sqrt{\pi_i k_{ij}/t}}.
\end{equation}
These normalized sources are orthonormal in the long term, satisfying $\lim_{t\rightarrow \infty} t \cdot \mathrm{Cov}[\hat{\lambda}_{ij},\hat{\lambda}_{ab}] = \delta_{ia}\delta_{jb}$.
Consequently, the covariance contribution of a specific transition $ij$ is simply the product of the projections of the observables onto this unit noise vector (via RIM):
\begin{align}
    \lim_{t\rightarrow \infty} t \cdot \mathrm{Cov}[x, \hat{\lambda}_{ij}] &= \lim_{t\rightarrow \infty} t \cdot \mathrm{Cov}[\sum_{ab} A_{x,ab}^{-1} \lambda_{ab}, \frac{\lambda_{ij}}{\sqrt{\pi_i k_{ij}/t}}] \nonumber \\
    &= \frac{1}{\sqrt{\pi_i k_{ij}/t}} (\pi_i k_{ij} A_{x,ij}^{-1}) \nonumber \\
    &= \sqrt{t \pi_i k_{ij}} ~ A_{x,ij}^{-1}.
\end{align}
Substituting this back into the covariance sum recovers the \textbf{Parseval-like identity} (Eq. 11 in the main text):
\begin{equation}
    \mathrm{Cov}[x,x'] \sim \sum_{ij} \mathrm{Cov}[x, \hat{\lambda}_{ij}] \cdot \mathrm{Cov}[x', \hat{\lambda}_{ij}].
\end{equation}
This confirms that the total fluctuation is the sum of squared sensitivities projected onto independent microscopic noise sources.

\section{Details of Result 1: Covariance Decomposition of Kinesin}

To generate the covariance landscapes in Fig. 2 of the main text, we utilized the 6-state network model for Kinesin developed by \citet{liepelt_kinesins_2007}. This model captures the chemomechanical coupling of the motor under varying external loads ($F$) and ATP concentrations ($[\text{ATP}]$).

\begin{figure}[h!]
    \centering
    \includegraphics[width=0.4\linewidth]{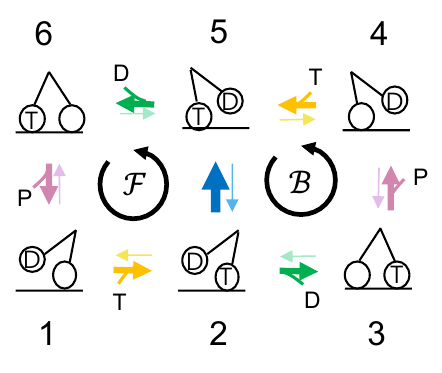}
    \caption{\textbf{Topology of the 6-state kinesin network.} The states are indexed $1-6$ based on the chemical composition of the trailing (left) and leading (right) heads ($\text{T}=\text{ATP}, \text{D}=\text{ADP}$). The network consists of a forward mechanical cycle $\mathcal{F}$ (left loop, physiological) and a backward hydrolysis cycle $\mathcal{B}$ (right loop). The central blue arrows denote the mechanical transitions ($2 \leftrightarrow 5$).}
    \label{fig:kinesin_topology_SI}
\end{figure}

\subsection{Model Topology and Transition Rates}
The model network consists of 6 chemical states (Fig. \ref{fig:kinesin_topology_SI}). The transitions form two primary cycles: the forward mechanical stepping cycle ($\mathcal{F}$) and the backward hydrolysis cycle ($\mathcal{B}$). For consistency with the provided Python code, we note that the code uses 0-based indexing (states 0-5), which corresponds directly to states 1-6 in this text (e.g., Code state 0 is State 1).\\

The transition rate from state $i$ to $j$ is parameterized as a function of the external load $F$ and chemical concentrations $[X]$:
\begin{equation}
    k_{ij}(F, [X]) = k_{ij}^0 \cdot \mathcal{I}_{ij}([X]) \cdot \Phi_{ij}(F).
\end{equation}
Here, $k_{ij}^0$ is the zero-load rate constant. The concentration factor $\mathcal{I}_{ij}([X])$ is $[X]$ for binding transitions and 1 otherwise. The force-dependence factors $\Phi_{ij}(F)$ are modeled as follows:\\

\paragraph{Mechanical Transitions ($2 \leftrightarrow 5$):}
These steps involve the physical translocation of the motor ($d=8$ nm) and follow a Bell-like exponential dependence:
\begin{align}
    \Phi_{25}(F) &= \exp\left(-\theta \frac{F d}{k_B T}\right), \quad \text{(Forward Step)} \\
    \Phi_{52}(F) &= \exp\left((1-\theta) \frac{F d}{k_B T}\right). \quad \text{(Backward Step)}
\end{align}
where $\theta$ is the load distribution factor.\\

\paragraph{Chemical Transitions ($i \leftrightarrow j$):}
Chemical transitions are modeled with a sigmoid-like dependence to account for the mechanical strain on the catalytic domain. This introduces the \textbf{load sensitivity factor} $\chi_{ij}$:
\begin{equation}
    \Phi_{ij}(F) = \frac{2}{1 + \exp\left(\chi_{ij} \frac{F d}{k_B T}\right)}.
\end{equation}
The parameter $\chi_{ij}$ characterizes the distance to the transition state in the chemical coordinate space under load.

\subsection{Physical Assumptions}
Based on the definitions above, the model from Liepelt and Lipowsky relies on the following physical assumptions which impose constraints on the parameters $k_{ij}^0$ and $\chi_{ij}$:
\begin{itemize}
    \item \textbf{Head Symmetry and Thermodynamic Constraints:} 
    Since the two motor heads are identical, the kinetics of chemical transitions are generally independent of the head's position. 
    Consequently, the model assigns identical zero-load rates to corresponding steps in the forward and backward cycles:
    \begin{align}
        k_{12}^0 = k_{45}^0 \quad (\text{ATP Binding}) &;~~~
        k_{56}^0 = k_{23}^0 \quad (\text{ADP Release}) \nonumber \\
        k_{65}^0 = k_{32}^0 \quad (\text{ADP Rebinding}) &;~~~
        k_{61}^0 = k_{34}^0 \quad (\text{Hydrolysis}) \nonumber \\
        k_{16}^0 = k_{43}^0 \quad (\text{Synthesis}) &
    \end{align}
    The only exception is the ATP unbinding rate $k_{54}^0$. 
    Unlike strict symmetry ($k_{54}^0 = k_{21}^0$), this rate is rigorously determined by the loop closure condition to ensure thermodynamic consistency.
    Fundamentally, the cycle affinity $\mathcal{A}$---defined kinetically as the log-ratio of the product of forward rates to backward rates around a closed loop---must match the physical thermodynamic free energy difference $\Delta G$ that serves as the ``battery'' of the cycle (e.g., $\Delta \mu_{\text{ATP}}$).
    Since the mechanical step introduces a strong kinetic asymmetry ($k_{25}^0 \neq k_{52}^0$), satisfying this potential condition across the coupled forward and backward cycles mandates a compensatory adjustment in the unbinding rate:
    \begin{equation}
        k_{54}^0 = k_{21}^0 \left( \frac{k_{52}^0}{k_{25}^0} \right)^2.
    \end{equation}
    This ensures that the model respects detailed balance at equilibrium and correctly describes the dissipation away from equilibrium.
    
    \item \textbf{No Mechanical Substeps:} The mechanical step occurs in a single transition ($2 \leftrightarrow 5$) without intermediate substeps.
    \item \textbf{Symmetric Load Dependence for Chemical Transitions:} The load dependence of chemical transitions is assumed to be symmetric, meaning the distance to the transition state is equal for forward and backward rates. This implies:
        $\chi_{ij} = \chi_{ji}.$
    Consequently, $\Phi_{ij}(F) = \Phi_{ji}(F)$, meaning the external load scales the forward and reverse chemical rates by the same factor.
    \item \textbf{Identical Kinetics for ADP Release and Hydrolysis:} To reduce the number of free parameters, the model assumes that the ADP release steps ($5\to 6, 2\to 3$) and the Hydrolysis steps ($6\to 1, 3\to 4$) share identical kinetic properties. This implies equality in both their zero-load rates and their load sensitivities ($k_{56}^0 = k_{61}^0$ and $\chi_{56} = \chi_{61}$). Consequently, these distinct chemical transitions proceed with identical rates under all load conditions:
    \begin{equation}
        k_{56}(F) = k_{61}(F) \quad \text{and} \quad k_{23}(F) = k_{34}(F).
    \end{equation}
\end{itemize}
\begin{table}[h]
\centering
\begin{tabular}{l | c | l}
\hline \hline
\textbf{Parameter} & \textbf{Value} & \textbf{Description} \\ \hline
$\theta$ & 0.30 & Load distribution factor \\
$k_{25}^0$  & $3.0 \times 10^5$ & Mechanical fwd rate \\
$k_{52}^0$  & 0.24 & Mechanical bwd rate \\
$k_{12}^0$  & 1.8 & ATP binding rate \\
$k_{56}^0$ & 200.0 & ADP release rate \\
$k_{61}^0$  & 200.0 & Hydrolysis rate \\
$\chi_{12}$  & 0.25 & Load factor (ATP) \\
$\chi_{56}$  & 0.05 & Load factor (ADP/Hydr) \\
\hline \hline
\end{tabular}
\caption{Parameter set used for the Kinesin model (Visscher et al.). Rates are in units of $s^{-1}$ (first order) or $\mu M^{-1} s^{-1}$ (second order). Only one rate is listed for symmetric pairs.}
\label{tab:kinesin_params}
\end{table}
\subsection{Parameter Sets}
We utilized the parameter set \citet{liepelt_kinesins_2007} fitted to the optical trap data from \citet{visscher_single_1999}. 
We selected this specific dataset because it provides comprehensive experimental measurements of the motor's \textbf{randomness parameter} (Fano factor---the variance normalized by the mean---of motor's velocity) under varying loads and ATP concentrations, which is essential for validating our fluctuation analysis.
The explicit values used in our calculations are listed in Table \ref{tab:kinesin_params}.

\subsection{The Randomness Parameter as the Fano factor of the mechanical flux}
Denote the stochastic motor displacement at time $t$ as $\Delta x(t) = d \cdot (N_+(t) - N_-(t))$, where $d=8$ nm is the step size, and $N_+(t)$ and $N_-(t)$ denote the cumulative counts of forward ($2\to 5$) and backward ($5\to 2$) mechanical steps up to time $t$.
We define the time-averaged net flux of the mechanical step as the random variable:
\begin{equation}
    \psi \equiv \frac{N_+(t) - N_-(t)}{t}.
\end{equation}
By the Law of Large Numbers, the time-averaged quantities converge to their ensemble means as $t \to \infty$:
\begin{align}
    \lim_{t\to\infty} d \cdot \psi  &= \langle v \rangle = d \cdot \langle \psi \rangle, \\
    \lim_{t\to\infty} \frac{\text{Var}(\Delta x(t))}{t} &= 2D = \lim_{t\to\infty} d^2 \cdot t \cdot \text{Var}(\psi).
\end{align}
The randomness parameter $r$ is defined as the Fano factor of the flux:
\begin{equation}
    r = \frac{2D}{\langle v \rangle d} = \lim_{t\rightarrow \infty} \frac{ t \cdot \text{Var}(\psi)}{\langle \psi \rangle}.
\end{equation}

\subsection{Covariance Decomposition Implementation}
To spatially resolve the sources of noise, we partitioned the 14 directed transitions of the network into 8 physically distinct groups, denoted here as the set $\mathcal{G}$. The total Fano factor $r$ of the motor velocity is the sum of the randomness contributions $S_g$ from each group:
\begin{equation}
    r = \sum_{g \in \mathcal{G}} S_g.
\end{equation}
Based on the RIM relation (Eq. 11 in the main text), the contribution $S_g$ is calculated by summing the squared-sensitivity projections of the individual edges $(i,j)$ belonging to that group:
\begin{equation}
    S_g = \frac{1}{\langle \psi \rangle} \sum_{(i,j) \in g} \frac{1}{\pi_i k_{ij}} \left( \frac{\partial \langle \psi \rangle}{\partial \ln k_{ij}} \right)^2 \sim \frac{1}{\langle \psi \rangle} \sum_{(i,j) \in g} \pi_i k_{ij} \left( [\mathbf{A}^{-1}]_{\psi, (ij)} \right)^2.
\end{equation}
The groups are defined to distinguish between the forward reaction steps (driving the cycles) and the reverse reaction steps (opposing the cycles).\\ 

Crucially, we aggregate the chemical transitions from both the forward and backward cycles into single categories.
This reflects the physical reality that corresponding chemical steps in both cycles are governed by the same environmental parameters (e.g., [ATP] or [ADP]), acting effectively as a single control knob.
Since an experimental perturbation to the concentration affects these rates simultaneously, summing their contributions provides a more physically meaningful measure of sensitivity than treating them in isolation. This aggregation also simplifies the covariance landscape for clearer visualization. The eight groups are:
\begin{itemize}
    \item \textbf{ATP Binding:} Aggregates the forward binding rates from both cycles ($0\to1, 3\to4$) vs. Release ($1\to0, 4\to3$).
    \item \textbf{Mechanical Step:} Forward ($1\to4$) vs. Backward ($4\to1$). 
    \item \textbf{ADP Release:} Aggregates the forward release rates from both cycles ($4\to5, 1\to2$) vs. Binding ($5\to4, 2\to1$).
    \item \textbf{Hydrolysis:} Aggregates the forward hydrolysis rates from both cycles ($5\to0, 2\to3$) vs. Synthesis ($0\to5, 3\to2$).
\end{itemize}
This decomposition allows us to identify whether the total randomness $r$ is dominated by the mechanical stepping itself or by the stochasticity of specific chemical steps.

\section{Details of Result 2: Scalable Gradient Evaluation}

In this section, we demonstrate how the algebraic closure of the inverse Jacobian matrix allows for computing the exact gradient of the randomness parameter with respect to \textit{all} rate constants in a single calculation step, avoiding expensive numerical perturbations.

\subsection{Exact Gradient via Algebraic Closure}
To apply the RIM formalism, we explicitly select the mechanical transition ($2 \leftrightarrow 5$) as the defining chord for the fundamental mechanical cycle. Consequently, the flux observable $\psi$ corresponds to a basis vector in the Jacobian $\mathbf{A}$. The covariance decomposition gives:
\begin{equation}
    \lim_{t\rightarrow \infty} t \cdot \text{Var}(\psi) = \sum_{(i,j)} \pi_i k_{ij} \left( [\mathbf{A}^{-1}]_{\psi,(ij)} \right)^2.
    \label{eq:2D_RIM}
\end{equation}
Thus, the randomness parameter is simply:
\begin{equation}
    r = \frac{1}{\langle \psi \rangle}\sum_{i\neq j} \pi_i k_{ij} \left( [\mathbf{A}^{-1}]_{\psi,(ij)} \right)^2.
\end{equation}
Our goal is to compute its gradient vector $\nabla_{\ln \mathbf{k}} r$.\\

The calculation is a straightforward calculus. The key step is differentiating the inverse matrix using the relation:
\begin{equation}
    \partial_{k_{ij}} \mathbf{A}^{-1} = - \mathbf{A}^{-1}~ (\partial_{k_{ij}} \mathbf{A})~\mathbf{A}^{-1}.
\end{equation}
This is derived by differentiating the identity $\mathbf{I}=\mathbf{A} \mathbf{A}^{-1}$.
Based on the structure of the Jacobian $\mathbf{A}$ (Fig. \ref{fig:general A construction}), where rows are indexed by edges (forces) and columns by observables (frequencies $f$ and fluxes), the derivative matrix $\partial_{k_{ij}} \mathbf{A}$ is extremely sparse. Its non-zero components appear \textbf{only} in the columns corresponding to node frequencies.\\

We can thus reduce the derivative relation into the following explicit component forms:
\begin{enumerate}
    \item \textbf{For a non-reference node ($i \neq m$):}
    The rate $k_{ij}$ appears linearly with a coefficient of $-1$ at the intersection of row $(ij)$ and column $f_i$. Thus, $\partial_{k_{ij}} \mathbf{A}$ has a single non-zero entry of $-1$:
    \begin{align}
        \partial_{k_{ij}} \mathbf{A}^{-1}_{\alpha,ab} &= -\sum_{cd,\beta} \mathbf{A}^{-1}_{\alpha,cd}~ (\partial_{k_{ij}} \mathbf{A})_{cd,\beta }~\mathbf{A}^{-1}_{\beta,ab},\nonumber \\
        &= -\sum_{cd,\beta} \mathbf{A}^{-1}_{\alpha,cd}~ (-\delta_{c,i}\delta_{d,j}~\delta_{\beta,f_i})~\mathbf{A}^{-1}_{\beta,ab},\nonumber \\
        &= + \mathbf{A}^{-1}_{\alpha,ij}~ \mathbf{A}^{-1}_{f_i,ab}.
    \end{align}
    \item \textbf{For the reference node ($i=m$):}
    Due to the normalization constraint $\sum f_n = 1$, the rate $k_{mj}$ appears with a coefficient of $+1$ in every column corresponding to an independent frequency $f_n$. Thus, the derivative is a sum over all basis nodes $n$:
    \begin{align}
        \partial_{k_{mj}} \mathbf{A}^{-1}_{\alpha,ab} &= -\sum_{cd,\beta} \mathbf{A}^{-1}_{\alpha,cd}~ (\partial_{k_{mj}} \mathbf{A})_{cd,\beta }~\mathbf{A}^{-1}_{\beta,ab},\nonumber \\
        &= -\sum_{cd,\beta} \mathbf{A}^{-1}_{\alpha,cd}~ (\delta_{c,m}\delta_{d,j} \sum_n \delta_{\beta,f_n})~\mathbf{A}^{-1}_{\beta,ab},\nonumber \\
        &=- \mathbf{A}^{-1}_{\alpha,mj}~ \sum_{n}\mathbf{A}^{-1}_{f_n,ab}.
    \end{align}
\end{enumerate}
Finally, since the derivatives of $\pi_i$ (mean frequency $\langle f_i \rangle$) and $\langle \psi \rangle$ can also be evaluated directly using $\mathbf{A}^{-1}$ (via RIM), the full gradient vector $\nabla r$ is evaluated efficiently in a single pass once $\mathbf{A}^{-1}$ is computed.

\subsection{Computational Complexity Analysis}
Here, we compare the theoretical scaling of our algebraic approach against the standard spectral deformation technique (Koza's method \cite{koza_general_1999}) for a general network with $N$ nodes and $M$ physical links. The number of rate parameters is $N_{\text{param}} = 2M$ (forward and backward rates).\\

To rigorously analyze the scaling crossover, we express the total computational time $\mathcal{T}$ as the sum of a one-time pre-calculation cost ($\mathcal{T}_{\text{setup}}$) and the cumulative cost of evaluating gradients for all parameters ($\mathcal{T}_{\text{iter}}$):
\begin{equation}
    \mathcal{T}_{\text{total}} = \mathcal{T}_{\text{setup}} + N_{\text{param}} \cdot \mathcal{T}_{\text{step}}.
\end{equation}

\subsubsection{Standard Spectral Method (Koza)}
The standard approach calculates the cumulants of a current via the generating function. One constructs a parameter-dependent generator matrix (or tilted matrix) $\tilde{\mathbf{W}}(s)$ of size $N \times N$, where the transition rates are modified by the conjugate variable $s$:
\begin{equation}
    [\tilde{\mathbf{W}}(s)]_{ij} = k_{ij} e^{s \cdot \theta_{ij}} - \delta_{i,j} \sum_{l} k_{il}.
\end{equation}
Here, $\theta_{ij} \in \{+1, -1, 0\}$ is the counting indicator for the transition $i\to j$.
For example, to compute the statistics of the net mechanical flux $\psi = (N_{25}-N_{52})/t$, one specifically sets the indicators for the mechanical transitions as $\theta_{25}=+1$ and $\theta_{52}=-1$, while setting all other $\theta_{ij}=0$.
The scaled variance is then obtained from the second derivative of the largest real eigenvalue $\lambda(s)$ of $\tilde{\mathbf{W}}(s)$ at $s=0$, $\lim_{t\rightarrow \infty} t ~\text{Var}[\psi]=\lambda''(0)$.\\

To compute the \textit{gradient} vector $\nabla_{\mathbf{k}} r$, one requires the sensitivity $\partial \lambda''(0) / \partial k_{uv}$ for every rate constant. Using finite differences, this necessitates re-evaluating the eigenvalue $\lambda(s)$ (and its numerical derivatives) for each perturbed parameter.
\begin{itemize}
    \item \textbf{Setup:} The cost is negligible ($\mathcal{T}_{\text{setup}} \approx 0$).
    \item \textbf{Step:} The bottleneck is the eigensolver. For a dense non-symmetric matrix, the cost scales as $\mathcal{T}_{\text{step}} \approx \mathcal{O}(N^3)$.
\end{itemize}
Substituting these into the total time yields:
\begin{equation}
    \mathcal{T}_{\text{Koza}} \approx 0 + (2M) ~\mathcal{O}(N^3) = \mathcal{O}(M \cdot N^3).
\end{equation}

\subsubsection{Algebraic RIM Method}
Our method shifts the computational burden to constructing and inverting the Jacobian matrix $\mathbf{A}$. The dimension of $\mathbf{A}$ is determined by the edge count, $D_{\mathbf{A}} = 2M$.
\begin{itemize}
    \item \textbf{Setup:} This is the bottleneck. Although $\mathbf{A}$ is sparse, its inverse $\mathbf{A}^{-1}$ is generally dense. Standard dense inversion scales cubically with the dimension: $\mathcal{T}_{\text{setup}} \approx \mathcal{O}(M^3)$.
    \item \textbf{Step:} Once $\mathbf{A}^{-1}$ is known, the gradient for each parameter involves multiplying specific matrix columns by the sparse derivative vector $\partial_{\mathbf{k}} \mathbf{A}$. Since $\partial_{\mathbf{k}} \mathbf{A}$ contains only $\mathcal{O}(1)$ non-zero elements, this vector-matrix product scales linearly with the matrix dimension: $\mathcal{T}_{\text{step}} \approx \mathcal{O}(M)$.
\end{itemize}
Substituting these into the total time yields:
\begin{equation}
    \mathcal{T}_{\text{RIM}} \approx \mathcal{O}(M^3) + (2M) \cdot \mathcal{O}(M) = \mathcal{O}(M^3) + \mathcal{O}(M^2).
\end{equation}
Note that the iterative term is sub-dominant ($\mathcal{O}(M^2)$) compared to the setup term ($\mathcal{O}(M^3)$).

\subsubsection{Scaling Comparison}
The relative performance depends on the network topology. We analyze two distinct limits:

\begin{enumerate}
    \item \textbf{Sparse / Physical Networks:}
    Biophysical models typically have localized connectivity, so $M \approx \mathcal{O}(N)$.
    \begin{align}
        \mathcal{T}_{\text{Koza}} &\sim (N) \cdot N^3 = \mathcal{O}(N^4) \\
        \mathcal{T}_{\text{RIM}}  &\sim (N)^3 + (N)^2 = \mathcal{O}(N^3)+\mathcal{O}(N^2)
    \end{align}
    RIM achieves an asymptotic speed-up of factor $N$. The overhead of inverting a larger matrix ($2N \times 2N$) is outweighed by avoiding $N$ repetitions of the eigensolver.
    
    \item \textbf{Dense  Networks ($M \approx \mathcal{O}(N^2)$:}
    In fully connected graphs, the number of edges grows quadratically.
    \begin{align}
        \mathcal{T}_{\text{Koza}} &\sim (N^2) \cdot N^3 = \mathcal{O}(N^5) \\
        \mathcal{T}_{\text{RIM}}  &\sim (N^2)^3 + (N^2)^2 = \mathcal{O}(N^6)+\mathcal{O}(N^4)
    \end{align}
    The scaling advantage inverts. For extremely dense graphs, Koza's method is asymptotically superior because the dimension of the RIM Jacobian ($N^2$) grows much faster than the spectral matrix ($N$).
\end{enumerate}

\paragraph{Note on Implementation.}
While $\mathbf{A}$ is a sparse matrix, its inverse $\mathbf{A}^{-1}$ is generally dense due to the global coupling of states in the NESS. Therefore, we utilize optimized dense linear algebra routines (e.g., LAPACK via NumPy) for the inversion. The scalability advantage of our method arises primarily from algebraic closure---computing the gradient requires only \textit{one} matrix inversion and matrix-vector multiplications---rather than from the use of sparse inversion algorithms. This avoids the $2M$ repeated eigensolver calls required by generating function methods.\\

\paragraph{Note on Empirical Scaling.} 
The complexity bounds derived above assume standard dense linear algebra scaling (e.g., cubic scaling for inversion and diagonalization). In practice, highly optimized numerical routines (such as LAPACK) often achieve better effective scaling (typically $\sim D^{2.4-2.8}$) for the intermediate matrix sizes models ($N \lesssim 100$). Consequently, while the \textit{relative} ranking of the two methods remains consistent with our derivation, the absolute empirical exponents observed in numerical benchmarks may be lower than the worst-case theoretical bounds.

\subsection{Numerical Benchmark: Extended Kinesin Model}
To empirically validate the scalability of our approach, we benchmarked both methods using a generalized Kinesin network model. This setup corresponds to the results presented in Fig. 3 of the main text and utilizes the provided Python code.

\subsubsection{Constructing the Extended Model}
We utilized a ``$\Theta$-shape'' topology (Fig. 3a in Main Text), which extends the standard 6-state Kinesin model \cite{liepelt_kinesins_2007} to arbitrary system sizes. The network consists of two fixed chemical branches (ATP-driven forward and backward cycles) connected by a central mechanical bridge.
To vary the state-space size $N_{\text{sys}}$, we subdivided the single mechanical step ($2 \leftrightarrow 5$ in the original model) into a linear chain of $N_{\text{sub}}$ mechanical substeps.
\begin{itemize}
    \item \textbf{Topology:} The system consists of $N_{\text{sys}} = 4 + (N_{\text{sub}} + 1)$ states. The number of edges scales linearly with the system size, $M \approx 3 N_{\text{sys}}$, representing a sparse physical topology typical of molecular motors.
    
    \item \textbf{Thermodynamic Consistency:} To ensure the model remains physically meaningful as $N_{\text{sub}}$ increases, we must preserve the total chemical potential difference (cycle affinity) across the entire loop. The generalized detailed balance condition requires that the product of forward rates divided by the product of backward rates around the cycle satisfies:
    \begin{equation}
        \underbrace{\left( \frac{k_{\text{mech}}^+}{k_{\text{mech}}^-} \right)}_{\text{Mechanical}} \cdot \underbrace{ \prod_{\text{chem}}\left( \frac{k_{\text{chem}}^+}{k_{\text{chem}}^-} \right)}_{\text{Chemical}} = \exp\left( \frac{\Delta \mu_{\text{ATP}} - F_{\text{load}} d}{k_B T} \right),
    \end{equation}
    where the chemical ratio includes contributions from ATP binding, hydrolysis, and ADP release.
    If the mechanical bridge is divided into $N_{\text{sub}}$ identical steps, the mechanical ratio becomes $(k_{\text{sub}}^+ / k_{\text{sub}}^-)^{N_{\text{sub}}}$. Therefore, to maintain the original driving force $\Delta \mu$, we follow Wagoner and Dill \cite{wagoner_mechanisms_2019} and scale the individual substep rates geometrically:
    \begin{equation}
        k_{\text{sub}}^\pm = \left( k_{\text{total}}^\pm \right)^{1/N_{\text{sub}}}.
    \end{equation}
    This ensures that the free energy drop per substep is exactly $\Delta G_{\text{total}} / N_{\text{sub}}$.
    
    \item \textbf{Numerical Stability:} To prevent artificial degeneracies in the linear solver arising from translational symmetry in the uniform chain, we applied a small random perturbation ($\pm 1\%$) to all transition rates.
\end{itemize}

\subsubsection{Scaling Results}
We computed the full gradient of the randomness parameter $r$ with respect to all edge rates for system sizes ranging from $N_{\text{sub}}=10$ to $320$. The computation times are plotted in Fig. 3b of the main text.

\begin{enumerate}
    \item \textbf{Koza's Method} (Red, $\approx N^{3.1}$):
    The brute-force spectral method requires iterating over all $2M$ parameters. For each parameter, it solves the eigenvalue problem for a sparse tilted matrix of size $N_{\text{sys}}$.
    The observed scaling exponent of $3.1$ is consistent with the theoretical prediction: it is the product of the number of parameters ($N_{\text{param}} \propto N$) and the effective cost of the sparse eigensolver ($\mathcal{T}_{\text{eig}} \propto N^{2.1}$). Note that while the theoretical worst-case for eigensolvers is $N^3$, optimized solvers (like ARPACK) typically achieve effective scaling closer to $N^{2.1}$ for banded matrices of this size.

    \item \textbf{Algebraic RIM Method} (Blue, $\approx N^{1.6}$):
    Our method performs a single matrix inversion of size $2M \times 2M$.
    The observed scaling exponent of $1.6$ is remarkably low. This indicates that for the intermediate matrix sizes benchmarked here ($D_{\mathbf{A}} \lesssim 1000$), the direct linear solver operates in a highly efficient pre-asymptotic regime, scaling significantly better than the theoretical cubic bound ($N^3$).
    Crucially, the scaling gap ($3.1 - 1.6 = 1.5$) is even larger than the conservative theoretical prediction of $1.0$ derived in the previous section. This effectively demonstrates that the ``setup overhead'' of RIM (inverting a larger matrix) is negligible compared to the ``iterative penalty'' of the spectral method.
\end{enumerate}

For a system size of $N_{\text{sys}} \approx 100$, our method achieves a speed-up of over two orders of magnitude ($>100\times$), confirming its practical utility for optimizing large-scale biophysical models.

\subsection{Numerical Benchmark: Dense All-to-All Networks}
To probe the performance limits of our edge-centric formalism, we benchmarked a fully connected network where every state is connected to every other state. This ``all-to-all'' topology ($M = N(N-1) \approx N^2$) represents the worst-case scenario for our approach, as the Jacobian dimension grows quadratically with the system size.

\begin{figure}[h]
    \centering
    \includegraphics[width=0.7\linewidth]{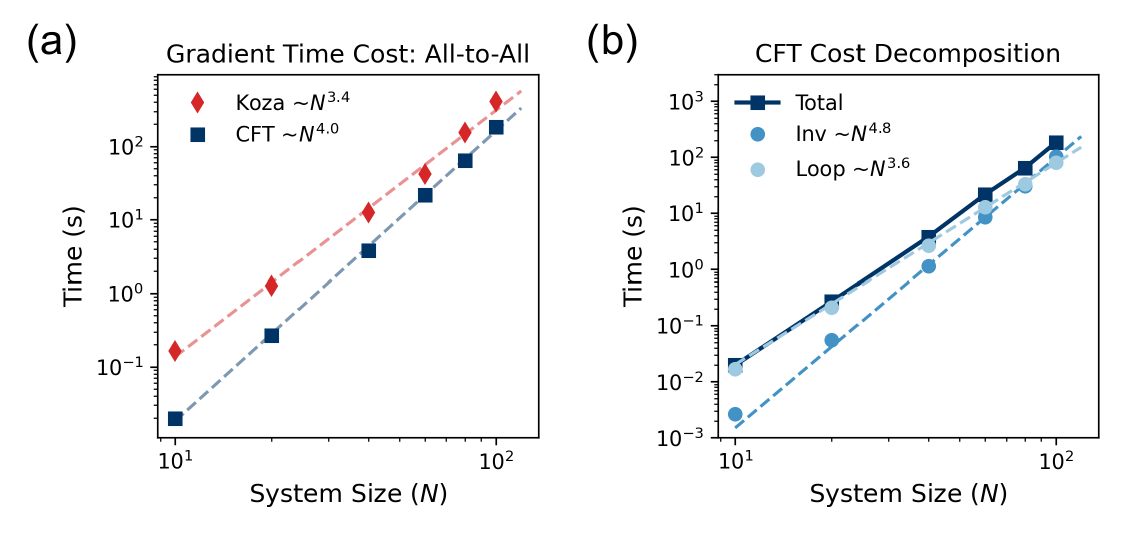}
    \caption{\textbf{Benchmarking on Dense All-to-All Networks.} \textbf{(a)} Comparison of total computation time between the Spectral method (Koza, red) and the Algebraic method (RIM, blue). Despite a lower asymptotic scaling exponent, Koza's method remains slower for $N \le 100$ due to large prefactors. \textbf{(b)} Component breakdown of the RIM method. For large $N$, the matrix inversion cost (Setup, medium blue) dominates with a scaling of $\approx N^{4.8}$, while the gradient assembly loop (Loop, light blue) scales as $\approx N^{3.6}$.}
    \label{fig: all-to-all}
\end{figure}

\subsubsection{Scaling Results}
We computed the gradient computation time for system sizes up to $N=100$. The results, shown in Fig. \ref{fig: all-to-all}a and decomposed in Fig. \ref{fig: all-to-all}b, reveal a performance crossover driven by the massive growth in connectivity.

\begin{enumerate}
    \item \textbf{Koza's Method} (Red, $\approx N^{3.4}$):
    The number of rate parameters grows quadratically as $N_{\text{param}} \approx 2N^2$. Consequently, the method requires solving the $N \times N$ eigenvalue problem $2N^2$ times.
    The observed scaling ($N^{3.4}$) is significantly better than the theoretical worst-case bound ($N^2 \times N^3 = N^5$). This suggests that for the small dense matrices examined here ($N \le 100$), the eigensolver operates in a pre-asymptotic regime with an effective scaling closer to $N^{1.4}$ per call. However, despite the favorable exponent, the absolute computation time is substantial due to the sheer number of repeated solver calls.

    \item \textbf{Algebraic RIM Method} (Blue, $\approx N^{4.0}$):
    The Jacobian matrix dimension expands to $2N^2 \times 2N^2$.
    The total time scales as $N^{4.0}$, which is notably faster than the theoretical dense inversion scaling of $(N^2)^3 = N^6$. Decomposing the cost (Fig. \ref{fig: all-to-all}b) clarifies this behavior:
    \begin{itemize}
        \item \textbf{Inverse Cost (Setup):} Scales as $\approx N^{4.8}$. Since the matrix dimension is $D \sim N^2$, this corresponds to an effective solver scaling of $D^{2.4}$, which is consistent with optimized dense linear algebra routines (e.g., LAPACK/BLAS) on modern hardware. For sufficiently large $N$, this term will eventually dominate.
        \item \textbf{Loop Cost (Gradient Assembly):} Scales as $\approx N^{3.6}$. This step involves iterating over $2N^2$ parameters and performing vector operations of size $2N^2$. The theoretical cost is $M^2 \sim N^4$, which closely matches the observed $N^{3.6}$.
    \end{itemize}
    In the benchmarked range ($N \le 100$), the Loop cost contributes significantly to the total time, weighing down the steeper scaling of the Inverse.
\end{enumerate}

\subsubsection{The Crossover Argument}
Although Koza's method possesses a superior asymptotic exponent ($3.4$ vs $4.0$) for dense graphs, \textbf{RIM remains computationally faster for system sizes up to $N \approx 100$}.
This crossover behavior arises from the distinct computational overheads of the two approaches:
\begin{itemize}
    \item \textbf{Koza (Cumulative Overhead):} This approach solves ``Many Small Problems.'' The total time is $2N^2 \times T_{\text{small}}$. The repeated overhead of function calls, memory allocation, and convergence checking for the iterative eigensolver creates a large cumulative prefactor.
    \item \textbf{RIM (Vectorization Efficiency):} This approach solves ``One Big Problem.'' The total time is $1 \times T_{\text{large}} + \text{fast loop}$. Modern CPUs are exceptionally efficient at handling single large matrix operations through cache optimization and vectorization.
\end{itemize}
Thus, even in the dense regime where our edge-based formalism is theoretically disadvantaged, the algebraic closure method remains a practical and competitive tool for exploring the design space of moderately sized, highly connected kinetic networks.

\section{Details of Result 3: Unifying Fluctuation-Response Relations}

In this section, we provide the detailed derivation of the three classes of response relations presented in Result 3 of the main text. We show that they are all mathematical consequences of the fundamental identity $\mathbf{A}^{-1}\mathbf{A} = \mathbf{I}$. Furthermore, we demonstrate how these relations unify and generalize and some recent results in the literature \citep{owen_universal_2020, aslyamov_nonequilibrium_2024, aslyamov_nonequilibrium_2025} by extending them to arbitrary observables and simplifying their algebraic bounds.

\begin{figure}[h]
    \centering
    \includegraphics[width=0.8\linewidth]{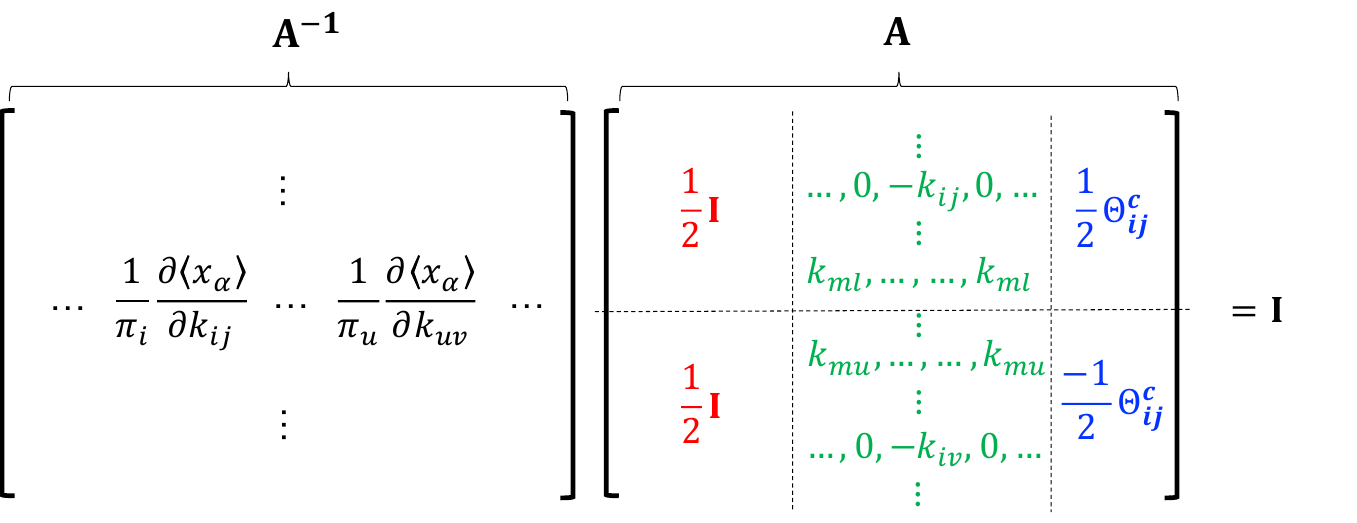}
    \caption{\textbf{Deriving Response Relations from Matrix Identity.} The Jacobian $\mathbf{A}$ (right) is structured into three blocks of columns. Multiplying the sensitivity vector (a row of $\mathbf{A}^{-1}$ associated with an arbitrary observable $x$) by these columns generates three distinct classes of constraints: Edge Reciprocity, Node Escaping Symmetry, and Cycle Symmetry.}
    \label{fig: AAinv_diagram}
\end{figure}

\subsection{Derivation of the Three Sets of Local Response Relations}

As illustrated in Fig. \ref{fig: AAinv_diagram}, the Jacobian matrix $\mathbf{A}$ is constructed using a specific basis corresponding to the fundamental observables: symmetric edge traffic $\phi_{ij}$ (with mean $\langle \phi_{ij} \rangle = \tau_{ij}$), node occupation $f_n$ (with mean $\langle f_n \rangle = \pi_n$), and cycle flux $\psi_c$ (with mean $\langle \psi_c \rangle = J_c$).
The fundamental identity $\mathbf{A}^{-1}\mathbf{A} = \mathbf{I}$ implies $\sum_\gamma [\mathbf{A}^{-1}]_{x, \gamma} [\mathbf{A}]_{\gamma, \beta} = \delta_{x, \beta}$. This generates three sets of constraints depending on the choice of the observable column $\beta$.

\subsubsection{Class I: Edge Reciprocity}
Selecting the column $\beta = \phi_{ij}$ corresponding to the symmetric traffic on edge $(ij)$, the identity yields:
\begin{equation}
    \frac{1}{\pi_i}\frac{\partial \langle x \rangle}{\partial k_{ij}} + \frac{1}{\pi_j}\frac{\partial \langle x \rangle}{\partial k_{ji}} = 2 \delta_{\langle x \rangle, \tau_{ij}}.
    \label{eq:class_I_reciprocity}
\end{equation}
This relation dictates that for any observable $x$ distinct from the local traffic $\tau_{ij}$ (e.g., a cycle flux $J_c$ or state probability $\pi_n$), the response to the forward rate and backward rate must be equal and opposite (weighted by $\pi$). This symmetry is the engine allowing for the simplification of kinetic bounds discussed below.

\subsubsection{Class II: Node Escaping Symmetry}
Selecting the column $\beta = f_n$ corresponding to the frequency of node $n$, the identity yields:
\begin{equation}
    \sum_{u (\neq m)} \frac{k_{mu}}{\pi_m}\frac{\partial \langle x \rangle}{\partial k_{mu}} - \sum_{l (\neq n)} \frac{k_{nl}}{\pi_n}\frac{\partial \langle x \rangle}{\partial k_{nl}} = \delta_{\langle x \rangle, \pi_n}.
    \label{eq:class_II_node}
\end{equation}
This describes the conservation of ``sensitivity flux'' at each node. It states that the sum of normalized sensitivities for rates entering a reference node $m$ differs from those leaving node $n$ only if the observable is the node occupancy $\pi_n$ itself.

\subsubsection{Class III: Cycle Symmetry}
Selecting the column $\beta = \psi_c$ corresponding to a fundamental cycle flux, the identity yields:
\begin{equation}
    \sum_{(ij) \in c^+} \frac{1}{\pi_i}\frac{\partial \langle x \rangle}{\partial k_{ij}} - \sum_{(ji) \in c^-} \frac{1}{\pi_j}\frac{\partial \langle x \rangle}{\partial k_{ji}} = 2 \delta_{\langle x \rangle, J_c}.
    \label{eq:class_III_cycle}
\end{equation}
This shows that the integrated sensitivity along a forward cycle must balance that of the backward cycle, with a net residue appearing only if the observable is the cycle flux $J_c$ itself.

\subsection{Generalizing \citet{owen_universal_2020} via Node Escaping Symmetry}
The ``universal stoichiometric response'' derived by \citet{owen_universal_2020} describes how state probabilities respond to perturbations. Their central result is:
\begin{align}
    \sum_{j(\neq i)}k_{ij}\frac{\partial\pi_{n}}{\partial k_{ij}} & =-\delta_{in}\pi_{n}(1-\pi_{n})+\left(1-\delta_{in}\right)\pi_{i}\pi_{n}.
    \label{eq:owen_target}
\end{align}
In this section, we show that this result is a specific instance of our \textbf{Node Escaping Symmetry}. 
Our framework is more general because the Node Escaping Symmetry applies to \textit{any} NESS observable (including  traffic and cycle fluxes), whereas Eq.~\eqref{eq:owen_target} is the specific projection of this symmetry onto state probabilities, constrained by normalization.

\subsubsection*{Step 1: The Node Energy Shorthand}
We first identify the physical meaning of the summation operator on the left-hand side of Eq.~\eqref{eq:owen_target}.
Consider a parameterization where each node has an ``energy'' $E_i$ such that $k_{ij} \propto e^{E_i}$. Physically, perturbing $E_i$ corresponds to changing the ``energy'' of state $i$ while fixing all other landscape dimensions (barriers and other state energies).
Mathematically, this provides a shorthand for the aggregate sensitivity:
\begin{equation}
    \frac{\partial \pi_n}{\partial E_i} \equiv \sum_{j(\neq i)} \frac{\partial \pi_n}{\partial k_{ij}} \frac{\partial k_{ij}}{\partial E_i} = \sum_{j(\neq i)} k_{ij} \frac{\partial \pi_n}{\partial k_{ij}}.
    \label{eq:energy_response_def}
\end{equation}

\subsubsection*{Step 2: Applying the Node Escaping Symmetry}
We now apply our Node Escaping Symmetry (Eq.~\ref{eq:class_II_node}) to the observable $\langle x \rangle = \pi_n$.
The symmetry relates the sensitivity flux originating from a reference node $m$ to that of an arbitrary node $i$:
\begin{equation}
    \sum_{u (\neq m)} \frac{k_{mu}}{\pi_m}\frac{\partial \pi_n}{\partial k_{mu}} - \sum_{j (\neq i)} \frac{k_{ij}}{\pi_i}\frac{\partial \pi_n}{\partial k_{ij}} = \delta_{in} - \delta_{mn}.
\end{equation}
Using shorthand $\partial_{E_i} \pi_n = \sum_{j\neq i} k_{ij} \partial_{k_{ij}} \pi_n$, we get:
\begin{equation}
    \frac{1}{\pi_m}\frac{\partial \pi_n}{\partial E_m} - \frac{1}{\pi_i}\frac{\partial \pi_n}{\partial E_i} = \delta_{in} - \delta_{mn}.
\end{equation}
To isolate the dependence on node $i$, we rearrange the terms, grouping all $i$-dependent terms on the left and $m$-dependent terms on the right:
\begin{equation}
    \frac{1}{\pi_i}\frac{\partial \pi_n}{\partial E_i} + \delta_{in} = \frac{1}{\pi_m}\frac{\partial \pi_n}{\partial E_m} + \delta_{mn}.
\end{equation}
Since the right-hand side depends only on the fixed reference node $m$ (and the target state $n$), while the left-hand side depends on the arbitrary source node $i$, both sides must be equal to a quantity $C_n$ that is \textit{constant} with respect to $i$. Thus, the normalized response must follow the structural form:
\begin{equation}
    \frac{1}{\pi_i}\frac{\partial \pi_n}{\partial E_i} = C_n - \delta_{in},
    \label{eq:structure_form}
\end{equation}
where $C_n$ is a constant determined by the reference term.

\subsubsection*{Step 3: Determining the Constant via Homogeneity}
To determine $C_n$, we invoke the time-scale invariance of Markov processes. Since steady-state probabilities $\pi_n$ depend only on rate ratios, they are homogeneous functions of degree 0 with respect to the rates.
This implies that a uniform shift in all node energies ($E_i \to E_i + \delta E$), which scales all rates globally, leaves $\pi_n$ unchanged:
\begin{equation}
    \sum_{\text{all } i} \frac{\partial \pi_n}{\partial E_i} = 0.
\end{equation}
Substituting our structural form $\partial_{E_i}\pi_n = \pi_i (C_n - \delta_{in})$ into this constraint:
\begin{align}
    \sum_i \pi_i (C_n - \delta_{in}) &= 0 \nonumber \\
    C_n \underbrace{\sum_i \pi_i}_{1} - \underbrace{\sum_i \pi_i \delta_{in}}_{\pi_n} &= 0 \quad \Rightarrow \quad C_n = \pi_n.
\end{align}

\subsubsection*{Step 4: Final Result}
With $C_n = \pi_n$, Eq.~\eqref{eq:structure_form} becomes:
\begin{equation}
    \frac{1}{\pi_i}\frac{\partial \pi_n}{\partial E_i} = \pi_n - \delta_{in}.
\end{equation}
Multiplying by $\pi_i$ and expanding the shorthand $\partial_{E_i}$, we recover:
\begin{equation}
    \sum_{j(\neq i)} k_{ij} \frac{\partial \pi_n}{\partial k_{ij}} = \pi_i \pi_n - \pi_i \delta_{in}.
\end{equation}
This matches Eq.~\eqref{eq:owen_target} exactly.\\

\textbf{Conclusion:} This derivation confirms that the stoichiometric response is not an isolated identity but a consequence of the Node Escaping Symmetry. Crucially, while Euler's homogeneity fixes the constant $C_n$ specifically for probabilities, the underlying symmetry structure (Eq.~\ref{eq:structure_form}) remains valid for \textit{any} observable, such as cycle fluxes or edge traffic. We next show how this leads to new response equalities.

\subsection{How Traffic and Cycle Flux response to escaping rate tuning.}
We derive here that for traffic $\tau_{ij}$ or cycle flux $J_c$ (or their linear combinations), their responses to tuning one node's escaping rate (i.e. $E_n$) have a simple expression: \begin{align}
    \frac{\partial \tau_{ij}}{\partial E_n} &= \sum_{l(\neq n)} k_{nl} \frac{\partial \tau_{ij}}{\partial k_{nl}}=\pi_n \tau_{ij};\\
    \frac{\partial J_c}{\partial E_n} &= \sum_{l(\neq n)} k_{nl} \frac{\partial J_c}{\partial k_{nl}} = \pi_n J_c;
\end{align}

\subsubsection*{Step 1: The Symmetry Condition}
For any flow observable $\langle y \rangle \notin \{\pi_n\}$ (where $\delta_{\langle y \rangle, \pi_n} = 0$), substituting the energy perturbation operator defined in Eq.~\eqref{eq:energy_response_def}, $\partial_{E_n} \equiv \sum_{j(\ne n)} k_{nj} \partial_{k_{nj}}$, the symmetry relation in Eq.~\eqref{eq:class_II_node} simplifies to:
\begin{equation}
    \frac{1}{\pi_m} \frac{\partial \langle y \rangle}{\partial E_m} - \frac{1}{\pi_n} \frac{\partial \langle y \rangle}{\partial E_n} = 0.
\end{equation}
This implies that the sensitivity, when normalized by the local population, is spatially uniform across the network:
\begin{equation}
    \frac{\partial \langle y \rangle}{\partial E_n} = C \pi_n,
    \label{eq:flux_scaling_structure}
\end{equation}
where $C$ is a constant specific to the observable $\langle y \rangle$ but independent of the node index $n$.

\subsubsection*{Step 2: Determining the Constant via Homogeneity}
To determine $C$, we again invoke the time-scale invariance. Unlike the probabilities $\pi$ (which are degree 0), fluxes and traffic rates have physical units of inverse time ($t^{-1}$). Consequently, they are homogeneous functions of \textit{degree 1} with respect to the transition rates.
A global shift in energy $E_n \to E_n + \delta E$ scales all rates by $e^{\delta E}$, and thus must scale the total magnitude of flow variables $\langle y \rangle$ by $e^{\delta E}$ without altering relative branching. Euler's homogeneous function theorem dictates:
\begin{equation}
    \sum_n \frac{\partial \langle y \rangle}{\partial E_n} = \langle y \rangle.
\end{equation}
Substituting the structural form from Eq.~\eqref{eq:flux_scaling_structure} into this constraint:
\begin{equation}
    \sum_n (C \pi_n) = C \underbrace{\sum_n \pi_n}_{1} = \langle y \rangle \quad \implies \quad C = \langle y \rangle.
\end{equation}

\subsubsection*{Step 3: The Scaling Law}
We thus arrive at the universal scaling relation for any steady-state flow observable $\langle y \rangle \in \{\tau_{ij}, J_c\}$:
\begin{equation}
    \frac{\partial \langle y \rangle}{\partial E_n} = \pi_n \langle y \rangle.
\end{equation}
This result proves that the perturbation of escaping rates (defined by tuning $E_n$) are degenerate controls for flux routing: they cannot alter the relative ratios of fluxes (pathways) in a network, but rather act locally as probability-weighted accelerators that scale the global magnitude of the flow.

\subsection{Barrier Tuning and the Net Flux Constraint}
We now consider the response to tuning ``the barrier height $B_{ij}$'' of a transition. Physically, raising a barrier decreases both the forward and backward rates simultaneously.
We define the barrier perturbation operator as:
\begin{equation}
    \frac{\partial}{\partial B_{ij}} \equiv -k_{ij}\frac{\partial}{\partial k_{ij}} - k_{ji}\frac{\partial}{\partial k_{ji}}.
\end{equation}
We show here that barrier tuning is governed by a strict Net Flux Constraint: it cannot affect any global observable $\langle x \rangle$ (except the local traffic $\tau_{ij}$) unless the edge carries a non-zero net flux ($J_{ij} \neq 0$).\\

For any observable $\langle x \rangle \neq \tau_{ij}$ (e.g., state probabilities $\pi_n$, cycle fluxes $J_c$, or other edge traffic $\tau_{uv}$, $uv \neq ij$), the right-hand side of the Edge Reciprocity relation (Eq.~\ref{eq:class_I_reciprocity}) vanishes:
\begin{equation}
    \frac{1}{\pi_i}\frac{\partial \langle x \rangle}{\partial k_{ij}} + \frac{1}{\pi_j}\frac{\partial \langle x \rangle}{\partial k_{ji}} = 0.
\end{equation}
This enforces a strict coupling between the forward and backward rate sensitivities:
\begin{equation}
    \frac{\partial \langle x \rangle}{\partial k_{ji}} = -\frac{\pi_j}{\pi_i} \frac{\partial \langle x \rangle}{\partial k_{ij}}.
\end{equation}
Substituting this into the barrier operator:
\begin{align}
    \frac{\partial \langle x \rangle}{\partial B_{ij}} &= -k_{ij}\frac{\partial \langle x \rangle}{\partial k_{ij}} - k_{ji}\left( -\frac{\pi_j}{\pi_i} \frac{\partial \langle x \rangle}{\partial k_{ij}} \right) \nonumber \\
    &= -\left( k_{ij} - k_{ji}\frac{\pi_j}{\pi_i} \right) \frac{\partial \langle x \rangle}{\partial k_{ij}} \nonumber \\
    &= -\frac{1}{\pi_i} \underbrace{(\pi_i k_{ij} - \pi_j k_{ji})}_{J_{ij}} \frac{\partial \langle x \rangle}{\partial k_{ij}}.
\end{align}
This yields the explicit response formula:
\begin{equation}
    \frac{\partial \langle x \rangle}{\partial B_{ij}} = -\frac{J_{ij}}{\pi_i} \frac{\partial \langle x \rangle}{\partial k_{ij}}.
    \label{eq:barrier_tuning_relation}
\end{equation}
\textbf{Conclusion:} If the edge is detailed-balanced ($J_{ij}=0$), the response vanishes ($\partial_{B_{ij}} \langle x \rangle = 0$).
This proves that barrier perturbations cannot tune $\langle x \rangle \neq \tau_{ij}$ if the edge is ``locally'' at equilibrium, $J_{ij}=0$. They acquire control authority over global landscape properties (like $\pi$ and $J_c$) only through the local net flux $J_{ij}$ generated by nonequilibrium driving.

\subsection{Generalizing the Flux Covariance from \citet{aslyamov_nonequilibrium_2025}}
\citet{aslyamov_nonequilibrium_2025} presented a decomposition for the covariance of cycle fluxes. We show this is a direct consequence of our Edge Reciprocity and covariance parsing.
The covariance decomposition (Result 1) is:
\begin{equation}
    t\cdot{\rm CoV}[\psi_{c},\psi_{c'}] = \sum_{i<j}\left(\frac{k_{ij}}{\pi_{i}}\frac{\partial J_{c}}{\partial k_{ij}}\frac{\partial J_{c'}}{\partial k_{ij}}+\frac{k_{ji}}{\pi_{j}}\frac{\partial J_{c}}{\partial k_{ji}}\frac{\partial J_{c'}}{\partial k_{ji}}\right). \label{eq: covariance decomposition of two net flux}
\end{equation}
Following \citet{owen_universal_2020} and \citet{aslyamov_nonequilibrium_2025}, one can consider the parameterization $k_{ij} = \exp [E_i-B_{ij}+F_{ij}/2]$ to consider various types of rate perturbations.

\subsubsection*{1. Barrier Sensitivity ($\partial B_{ij}$)}
Perturbing the symmetric barrier parameter $B_{ij}$ leads to $\frac{\partial}{\partial B_{ij}} \coloneqq -k_{ij}\frac{\partial}{\partial k_{ij}} - k_{ji}\frac{\partial}{\partial k_{ji}}$.
Since $J_c$ is not a local traffic variable, Eq. \eqref{eq:class_I_reciprocity} implies the reciprocity:
\begin{equation}
    \frac{1}{\pi_{i}}\frac{\partial J_c}{\partial k_{ij}} = -\frac{1}{\pi_{j}}\frac{\partial J_c}{\partial k_{ji}}.
    \label{eq:reciprocity_flux}
\end{equation}
This allows us to equate the rate sensitivity solely to the barrier sensitivity:
\begin{equation}
    \frac{\partial J_c}{\partial B_{ij}} = -\left(k_{ij} - k_{ji}\frac{\pi_j}{\pi_i}\right) \frac{\partial J_c}{\partial k_{ij}} = -\frac{J_{ij}}{\pi_i} \frac{\partial J_c}{\partial k_{ij}} \quad \Rightarrow \quad \frac{\partial J_c}{\partial k_{ij}} = -\frac{\pi_i}{J_{ij}} \frac{\partial J_c}{\partial B_{ij}}.\label{eq: dJ/dB = ... dJ/dk}
\end{equation}
Now, we substitute the reciprocity relation Eq.~\eqref{eq:reciprocity_flux} back into the covariance sum in Eq. \eqref{eq: covariance decomposition of two net flux}. The term inside the parenthesis for a given edge $ij$ becomes:
\begin{align}
    &\frac{k_{ij}}{\pi_{i}}\frac{\partial J_{c}}{\partial k_{ij}}\frac{\partial J_{c'}}{\partial k_{ij}}+\frac{k_{ji}}{\pi_{j}}\left(-\frac{\pi_{j}}{\pi_{i}}\frac{\partial J_{c}}{\partial k_{ij}}\right)\left(-\frac{\pi_{j}}{\pi_{i}}\frac{\partial J_{c'}}{\partial k_{ij}}\right) \nonumber \\
    =\;& \left(\frac{k_{ij}}{\pi_{i}} + \frac{k_{ji}\pi_j}{\pi_i^2} \right) \frac{\partial J_{c}}{\partial k_{ij}}\frac{\partial J_{c'}}{\partial k_{ij}} \nonumber \\
    =\;& \frac{1}{\pi_{i}^{2}}\underbrace{(\pi_{i}k_{ij}+\pi_{j}k_{ji})}_{\tau_{ij}}\frac{\partial J_{c}}{\partial k_{ij}}\frac{\partial J_{c'}}{\partial k_{ij}}. \label{eq:traffic_form_term}
\end{align}
Finally, plugging in the barrier sensitivity from Eq.~\eqref{eq: dJ/dB = ... dJ/dk}, we obtain:
\begin{align}
    \dots &= \frac{\tau_{ij}}{\pi_{i}^{2}}\left(-\frac{\pi_{i}}{J_{ij}}\frac{\partial J_{c}}{\partial B_{ij}}\right)\left(-\frac{\pi_{i}}{J_{ij}}\frac{\partial J_{c'}}{\partial B_{ij}}\right) \nonumber \\
    &= \frac{\tau_{ij}}{J_{ij}^{2}}\frac{\partial J_{c}}{\partial B_{ij}}\frac{\partial J_{c'}}{\partial B_{ij}}.
\end{align}
Summing over all edges $i<j$ recovers the first result from \citet{aslyamov_nonequilibrium_2025}:
\begin{align}
    t\cdot{\rm CoV}[\psi_{c},\psi_{c'}] = \sum_{i<j} \frac{\tau_{ij}}{J_{ij}^2} \frac{\partial J_c}{\partial B_{ij}} \frac{\partial J_{c'}}{\partial B_{ij}}.
\end{align}

\subsubsection*{2. Force Sensitivity ($\partial F_{ij}$)}
In \citet{aslyamov_nonequilibrium_2025}, the covariance is also expanded by the responses to the driving force $F_{ij}$, defined by $\partial_{F_{ij}}\coloneqq \frac{1}{2}\left ( k_{ij} \partial_{k_{ij}} - k_{ji} \partial_{k_{ji}} \right)$.
Using the reciprocity relation in Eq.~\eqref{eq:reciprocity_flux} to express $\partial_{k_{ji}}$ in terms of $\partial_{k_{ij}}$, the force sensitivity simplifies to:
\begin{align}
    \frac{\partial J_c}{\partial F_{ij}} 
    &= \frac{1}{2}\left( k_{ij}\frac{\partial J_c}{\partial k_{ij}} - k_{ji}\left(-\frac{\pi_j}{\pi_i}\frac{\partial J_c}{\partial k_{ij}}\right) \right) \nonumber \\
    &= \frac{1}{2\pi_i} \underbrace{(\pi_i k_{ij} + \pi_j k_{ji})}_{\tau_{ij}} \frac{\partial J_c}{\partial k_{ij}} 
    = \frac{\tau_{ij}}{2\pi_i} \frac{\partial J_c}{\partial k_{ij}}.
\end{align}
Inverting this relation gives the rate sensitivity explicitly in terms of the force sensitivity:
\begin{equation}
    \frac{\partial J_c}{\partial k_{ij}} = \frac{2\pi_i}{\tau_{ij}} \frac{\partial J_c}{\partial F_{ij}}.
\end{equation}
We substitute this back into the traffic-form of the covariance sum derived in Eq.~\eqref{eq:traffic_form_term}. The term for edge $ij$ becomes:
\begin{align}
    \frac{\tau_{ij}}{\pi_i^2} \left( \frac{\partial J_c}{\partial k_{ij}} \right) \left( \frac{\partial J_{c'}}{\partial k_{ij}} \right)
    &= \frac{\tau_{ij}}{\pi_i^2} \left( \frac{2\pi_i}{\tau_{ij}} \frac{\partial J_c}{\partial F_{ij}} \right) \left( \frac{2\pi_i}{\tau_{ij}} \frac{\partial J_{c'}}{\partial F_{ij}} \right) \nonumber \\
    &= \frac{\tau_{ij}}{\pi_i^2} \left( \frac{4\pi_i^2}{\tau_{ij}^2} \right) \frac{\partial J_c}{\partial F_{ij}} \frac{\partial J_{c'}}{\partial F_{ij}} \nonumber \\
    &= \frac{4}{\tau_{ij}} \frac{\partial J_c}{\partial F_{ij}} \frac{\partial J_{c'}}{\partial F_{ij}}.
\end{align}
Summing over all edges recovers the complementary expansion in \citet{aslyamov_nonequilibrium_2025}:
\begin{align}
    t\cdot{\rm CoV}[\psi_{c},\psi_{c'}] = \sum_{i<j} \frac{4}{\tau_{ij}} \frac{\partial J_c}{\partial F_{ij}} \frac{\partial J_{c'}}{\partial F_{ij}}.
\end{align}

\subsubsection*{Summary: Unifying the Decompositions}
This derivation establishes that the flux covariance decompositions derived in Ref.~\cite{aslyamov_nonequilibrium_2025} are {specific instances} of our general covariance parsing (Result 1).
Our framework provides the overarching geometry: by restricting the general observables $x$ and $x'$ to cycle fluxes ($J_c, J_{c'}$) and applying our \textbf{Edge Reciprocity} to map the canonical rate sensitivities ($\partial_{k}$) onto the specific barrier ($\partial_{B}$) and force ($\partial_{F}$) parameterizations, we exactly recover their identities.
Consequently, our $\mathbf{A}^{-1}$ formalism generalizes these fluctuation-response dualities beyond net fluxes, making them applicable to the full set of nonequilibrium observables including traffic and state occupancies.

\subsection{Translating the Kinetic Bounds from \citet{aslyamov_nonequilibrium_2024}}
\citet{aslyamov_nonequilibrium_2024} derived bounds on the response to the parameter $B_{ij}$. They proved that the quantities $\Delta_{ij} \equiv 1 + \partial_{B_{ij}} \ln |J_{ij}|$ and $\nabla_{ij} \equiv (\tau_{ij}/J_{ij})(1 + \partial_{B_{ij}} \ln \tau_{ij})$ satisfy:
\begin{align}
    0 \le \Delta_{ij} \le 1 \quad \text{and} \quad |\nabla_{ij}| \le \Delta_{ij}.
\end{align}
Our Edge Reciprocity translates these directly into simple inequalities for elementary rate sensitivities.

\subsubsection*{Step 1: Expressing $\Delta_{ij}$.}
Using the definition and the flux reciprocity (Eq. \ref{eq: dJ/dB = ... dJ/dk}):
\begin{align}
    \Delta_{ij} &= 1 + \frac{1}{J_{ij}} \frac{\partial J_{ij}}{\partial B_{ij}} \nonumber \\
    &= 1 + \frac{1}{J_{ij}} \left( -\frac{J_{ij}}{\pi_i} \frac{\partial J_{ij}}{\partial k_{ij}} \right) \nonumber \\
    &= 1 - \frac{1}{\pi_i} \frac{\partial J_{ij}}{\partial k_{ij}}.
\end{align}

\subsubsection*{Step 2: Expressing $\nabla_{ij}$.}
We apply the Edge Reciprocity to the traffic observable $\tau_{ij}$. Since $\tau_{ij}$ has a non-zero self-response ($2\delta=2$), Eq. \eqref{eq:class_I_reciprocity} gives $\frac{1}{\pi_i}\frac{\partial \tau_{ij}}{\partial k_{ij}} + \frac{1}{\pi_j}\frac{\partial \tau_{ij}}{\partial k_{ji}} = 2$.
The sensitivity to $B$-perturbation is:
\begin{align}
    \frac{\partial \tau_{ij}}{\partial B_{ij}} &= -k_{ij}\frac{\partial \tau_{ij}}{\partial k_{ij}} - k_{ji}\frac{\partial \tau_{ij}}{\partial k_{ji}} \nonumber \\
    &= -k_{ij}\frac{\partial \tau_{ij}}{\partial k_{ij}} - k_{ji} \pi_j \left( \frac{2}{\pi_j} - \frac{1}{\pi_i}\frac{\partial \tau_{ij}}{\partial k_{ij}} \right) \nonumber \\
    &= - \left( k_{ij} - \frac{\pi_j k_{ji}}{\pi_i} \right) \frac{\partial \tau_{ij}}{\partial k_{ij}} - 2 k_{ji} \pi_j \nonumber \\
    &= -\frac{J_{ij}}{\pi_i} \frac{\partial \tau_{ij}}{\partial k_{ij}} - (\tau_{ij} - J_{ij}).
\end{align}
Substituting this into the definition of $\nabla_{ij}$:
\begin{align}
    \nabla_{ij} &= \frac{\tau_{ij}}{J_{ij}} \left( 1 + \frac{1}{\tau_{ij}} \frac{\partial \tau_{ij}}{\partial B_{ij}} \right) \nonumber \\
    &= \frac{\tau_{ij}}{J_{ij}} + \frac{1}{J_{ij}} \left( -\frac{J_{ij}}{\pi_i} \frac{\partial \tau_{ij}}{\partial k_{ij}} - \tau_{ij} + J_{ij} \right) \nonumber \\
    &= 1 - \frac{1}{\pi_i} \frac{\partial \tau_{ij}}{\partial k_{ij}}.
\end{align}

\subsubsection*{Step 3: Deriving the Hierarchy.}
Substituting these expressions back into the Aslyamov bounds allows us to derive the kinetic hierarchy. We utilize the linear relations between CFT observables and one-way fluxes: $\tau_{ij} = p_{ij} + p_{ji}$ and $J_{ij} = p_{ij} - p_{ji}$. The derivatives satisfy:
\begin{equation}
    \frac{\partial \tau_{ij}}{\partial k_{ij}} = \frac{\partial p_{ij}}{\partial k_{ij}} + \frac{\partial p_{ji}}{\partial k_{ij}} \quad \text{and} \quad \frac{\partial J_{ij}}{\partial k_{ij}} = \frac{\partial p_{ij}}{\partial k_{ij}} - \frac{\partial p_{ji}}{\partial k_{ij}}.
\end{equation}
We now expand the three inequalities implied by $|\nabla_{ij}| \le \Delta_{ij}$ and $0 \le \Delta_{ij} \le 1$:
\begin{enumerate}
    \item The Le Chatelier Bound ($\nabla_{ij} \le \Delta_{ij}$):
\begin{align}
    1 - \frac{1}{\pi_i} \frac{\partial \tau_{ij}}{\partial k_{ij}} &\le 1 - \frac{1}{\pi_i} \frac{\partial J_{ij}}{\partial k_{ij}} \nonumber \\
    -\frac{\partial \tau_{ij}}{\partial k_{ij}} &\le -\frac{\partial J_{ij}}{\partial k_{ij}} \implies \frac{\partial \tau_{ij}}{\partial k_{ij}} \ge \frac{\partial J_{ij}}{\partial k_{ij}}.
\end{align}
Substituting the $p$-derivatives:
\begin{align}
    \left( \frac{\partial p_{ij}}{\partial k_{ij}} + \frac{\partial p_{ji}}{\partial k_{ij}} \right) &\ge \left( \frac{\partial p_{ij}}{\partial k_{ij}} - \frac{\partial p_{ji}}{\partial k_{ij}} \right) \nonumber \\
    2 \frac{\partial p_{ji}}{\partial k_{ij}} &\ge 0 \quad \implies \quad \frac{\partial p_{ji}}{\partial k_{ij}} \ge 0.
\end{align}
    \item The Causality Bound ($\Delta_{ij} \le 1$):
\begin{align}
    1 - \frac{1}{\pi_i} \frac{\partial J_{ij}}{\partial k_{ij}} &\le 1 \implies \frac{\partial J_{ij}}{\partial k_{ij}} \ge 0.
\end{align}
In terms of one-way fluxes, this yields:
\begin{equation}
    \frac{\partial p_{ij}}{\partial k_{ij}} - \frac{\partial p_{ji}}{\partial k_{ij}} \ge 0 \quad \implies \quad \frac{\partial p_{ij}}{\partial k_{ij}} \ge \frac{\partial p_{ji}}{\partial k_{ij}}.
\end{equation}
\item The Population Depletion Bound ($-\Delta_{ij} \le \nabla_{ij}$):
\begin{align}
    -\left( 1 - \frac{1}{\pi_i} \frac{\partial J_{ij}}{\partial k_{ij}} \right) &\le 1 - \frac{1}{\pi_i} \frac{\partial \tau_{ij}}{\partial k_{ij}} \nonumber \\
    -1 + \frac{1}{\pi_i} \frac{\partial J_{ij}}{\partial k_{ij}} &\le 1 - \frac{1}{\pi_i} \frac{\partial \tau_{ij}}{\partial k_{ij}} \nonumber \\
    \frac{1}{\pi_i} \left( \frac{\partial J_{ij}}{\partial k_{ij}} + \frac{\partial \tau_{ij}}{\partial k_{ij}} \right) &\le 2.
\end{align}
Substituting the sum $\partial_k J + \partial_k \tau = 2 \partial_k p_{ij}$:
\begin{align}
    \frac{1}{\pi_i} \left( 2 \frac{\partial p_{ij}}{\partial k_{ij}} \right) &\le 2 \quad \implies \quad \frac{\partial p_{ij}}{\partial k_{ij}} \le \pi_i.
\end{align}
\end{enumerate}
Combining these three results yields the strict kinetic hierarchy presented in Eq. (5) of the Main Text:
\begin{equation}
    \pi_{i} \ge \frac{\partial p_{ij}}{\partial k_{ij}} \ge \frac{\partial p_{ji}}{\partial k_{ij}} \ge 0.
\end{equation}

\end{document}